# Comments to support the Dipole Dynamical Model (DDM) of Ball Lightning (BL)[1]

V. N. Soshnikov[2]

Plasma Physics Dept.,

All-Russian Institute of Scientific and Technical Information
of the Russian Academy of Sciences
(VINITI, Usievitcha 20, 125315 Moscow, Russia)

## Abstract

I present estimates to justify previously proposed by me heuristic Dipole Dynamical Model (DDM) of Ball Lightning (BL). The movement and energy supplying to the dipole BL are due to the atmospheric electric field. Crucial for the detailed analysis of BL is using the new relation of balance of the force of atmospheric electric field (*per unit mass of electron cloud*) and dipole forces electrons-ions within BL dipole (*per unit mass of BL*) as the first necessary condition for the existence of BL as an integer. This model is unique because, unlike existing static models, fundamental condition for the existence of Ball Lightning is its forward motion. The virial theorem limiting BL power does not apply to BL which is not closed system like the Sun or Galaxy systems and is strongly dependent part of the infinitely extended in time and space large system. Stability of BL is due to two free parameters with the fundamental role of thermodynamic non-equilibrium, ionization, recombination and translational movement with energy loss by radiation and also excess volumetric positive charge. Stability of BL is not related to the presence of any external shells. Polarization degree of BL plasma is characterized by polarizability factor $\gamma$. An example is presented of calculating the stability of option BL. There is also a possible connection of stability BL with statistical distributions of the atmospheric electric field in time and space. Destruction of BL can also occur due to arising kinematical instability with its accelerating (or decelerating) movement. Maximal energy density in BL DDM does not exceed the value $E_{spec} < (10^8 - 10^9)$ J/m$^3$. Resulting indefinitely long BL lifetime is also discussed. BL has no outer shell and no any rigid or elastic inner microstructure elements.

## 1. Introduction

The variety of manifestations of BL leads to the conclusion that the term BL is the collective name for similar but possible different physical phenomena, including those described by the cluster model, the formation and fragmentation of vortex rings in the vicinity of the parent channel of linear lightning, the pinch effect (the beaded lightning) etc.

Estimates of the energy density in observed BL lead in dependence on BL size up to $(2-3)$ m and more (if exist) to the values up to $\sim 10^{12}$ J/m$^3$ (see [1], [2]). To explain such enormous energies and energy densities there are offered many primary exotic and even fantastic models (see [3] and detailed extensive review [1] and [4]): BL may be a black hole; it may be possible the presence in BL nuclear reactions with the release of nuclear energy; the existence of antimatter in the BL; BL as a set of Rydberg atoms; BL as a closed circular cavity filled with light energy; BL is due to the phase transition to supercooled (300 K) nonideal plasma [5a, 5b]; more realistic and robust, energy source of BL is collective electron-ion oscillations [6] (that may be applicable to fixed (static) plasmoid, but not to typical movable BL); very sophisticated quantum-mechanical constructions of G. C. Dijkhius (see in [1]) and many other such kind hypotheses and models with a possible explanation for the only some selected observed features of BL. Several models associate BL energy with its borrowing from linear lightning (see [1], [2]): there are models of the dynamic MHD capacitors, electrostatic microcapacitors, chemical models of combustion of diluted atomic clusters of Si, which were formed during the impact of a linear lightning into the ground. The main drawback of these models is the inability to explain the sometimes observed enormous energy densities of BL and its lifetime. However, detailed analysis of dozens of different models of BL is beyond the scope of this article.

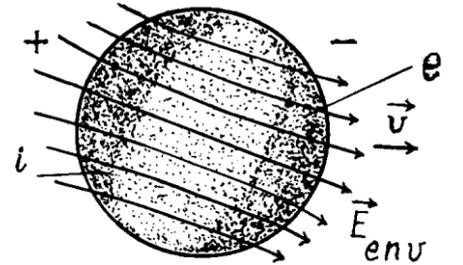

*Fig. 1. Gas-plasma dipole.*

In all these cases out of the question is the large energy of the electrostatic charge separation with a strong polarization of the plasma cloud BL by atmospheric electric field to form a plasma cloud dipole.

There is reported that the detailed dipole model of BL [7, 8] explains most of the observed properties of BL. For the first time is revealed the fundamental role of energy loss by radiation as the main source of losses, determining the necessity of energy supplying movement of BL along force lines of atmospheric electric field and playing stabilizing role in the relatively long lifetime of BL.

Described previously dipole dynamical model of BL [7] is based on an assumption of its independence (no direct relation with the linear lightning) because energy supporting BL is due to the atmospheric electric field, which is an integral part of this amazing atmospheric phenomenon. This particularly may indicate that the strong lethality and the devastating effect of BL can be due not only to BL itself, but possibly to discharge of its attendant strong atmospheric electric field, an indicator and conductor of which it is.

## 2. Problem setting and estimates [3, 4]

The DDM is based on two simple ideas: the observed movement of BL is due to the atmospheric electric field $E_{env}$ and at dipole nature of BL its movement is due to the influence of this field on electrons and compensation of $E_{env}$ field action on ions

| Table 1. Relative momentum loss $\tau$ * | | | | | | | |
|---|---|---|---|---|---|---|---|
| $T_e$, K | $T_i$, K | $r_{FB}$, cm | $\gamma E_{env}$, V/cm | $\alpha$ | $n_e^{(FB)}$, cm$^{-3}$ $\times 10^{-15}$ | $v_{FB}$, m/s | $\tau$ |
| 8500 | 400 | 5 | 162 | 8.4 | 10.1 | $1.73 \times 10^{-2}$ | 1.32 |
| 8500 | 3000 | 5 | 9.7 | 9.1 | 0.93 | 2.53 | $2.9 \times 10^{-4}$ |
| 10 000 | 400 | 5 | 343 | 20 | 21 | $3.4 \times 10^{-3}$ | $2.8 \times 10^{-2}$ |
| 10 000 | 1000 | 10 | 46.6 | 49.3 | 3.47 | $1.85 \times 10^{-2}$ | $7.0 \times 10^{-3}$ |
| 10 000 | 3000 | 5 | 52.8 | 52.6 | 2.97 | 0.26 | $1.56 \times 10^{-3}$ |
| 12 000 | 1000 | 10 | 122 | 136 | 5.61 | $4.4 \times 10^{-3}$ | $6.3 \times 10^{-3}$ |
| 12 000 | 3000 | 10 | 38.6 | 136 | 1.87 | $4.13 \times 10^{-2}$ | $1.14 \times 10^{-3}$ |

* $n_e^{(FB)} = (n_e [9] / \alpha) \ll n_m$; $\alpha$ is degree of non-equilibrium (see text below) and $n_e^{(FB)} \sim n_m x_e [9] / \alpha$, $\alpha \geq 1$. At $T_i = T_m = T_e \lesssim 12\,500$ K ion are dominated by only single ions; in the region $T_m = T_e \lesssim (10\,000 - 12\,000)$ K there prevail ions $O^+ \div O_2^+, N^+$. In all tables without account for energy loss to supporting the more weak forward recoil jet (see also below Section 6).

due to the recoil transfer of ions momentum to the cold dense air molecules in the back of BL with creation recoil BL jet. The energy of the atmospheric electric field $|e|\gamma E_{env} N_e s$, where $N_e$ is electron content, $\gamma$ is polarization factor, $s$ is travel distance of BL, is mostly transmitted to electrons and then spent to radiation but at the same extent to creation forward recoil jet which create braking resistance to gas-plasma BL movement. Gasdynamic resistance to movement BL, if to consider it as a firm body, is negligibly small [7]. Really gasdynamic resistance of BL to movement is created mainly by radiation energy losses.

Note, that the connection of BL with statistics of distributions of the environment electric field is still not investigated. It is also shown below that due to lateral ambipolar diffusion, BL has a significant positive charge.

It is shown [7] that BL confines neutral molecules and moves as an integer due to large ion collision recharge cross sections and momentum transfer due to Coulomb attraction and predominantly inelastic collisions of electrons, which are kept by positive charge, with neutral molecules. Thus, BL is inhomogeneous gas-plasma ball, although this heterogeneity and difference from the ideal spherical shape can be visually unobtrusive (Fig.1).

The cloud of electrons in an external electric field plays the role of locomotive for simultaneous forward movement BL with the cloud of ions and neutrals. This assumption is confirmed by comparing the total flux of ions onto the rear boundary of BL $n_i \bar{v}_x^i (T_i) / 2$ where $n_i = n_e$ is density of ions in the fireball (FB), $\bar{v}_x^i (T_i)$ is average $x$-component of velocity of ion flow normal to the rear bound of BL, with a diffusion flux $\mu_i n_i \gamma E_{env}$ of "runaway ions" where ion mobility $\mu_i$ is tightly related with the ion diffusion coefficient $D_i$. Using elementary relations

$$\mu_i \simeq \frac{|e|}{kT_{env}} \frac{\lambda \bar{v}_{rel}(T_{env})}{3}, \quad \bar{v}_{rel} = \frac{2}{\sqrt{\pi}} \sqrt{\frac{kT_{env}}{m_i}}, \quad \lambda \simeq \frac{1}{n_L \sigma_{coll}}, \quad \bar{v}_x^i (T_i) = \int_0^\infty v_x e^{-m_i v_x^2/2kT_i} dv_x \Big/ \int_0^\infty e^{-m_i v_x^2/2kT_i} dv_x = \sqrt{\frac{2kT_i}{\pi m_i}} \qquad (1)$$

where $n_L = 2.5 \times 10^{19} \text{cm}^{-3}$ (Loschmidt number at 20 $^{O}$C), elastic collision cross section $\sigma_{coll} \sim 2.5 \times 10^{-15} \text{cm}^2$, we obtain for the flux ratio $\tau$ (with the notations [7])

$$\tau \simeq \frac{\sqrt{2}}{3} \frac{|e|\gamma E_{env}}{k\sqrt{T_{env} T_i}} \frac{1}{n_L \sigma_{coll}} \simeq 2.78 \times 10^{-2} \frac{\gamma E_{env}}{\sqrt{T_{env} T_i}}, \qquad (2)$$

where environmental field $E_{env}$ is in V/cm, $T_i$, K is ion temperature in the fireball (FB), $T_{env} = 293$K, $\tau$ is relative loss of recoil momentum of ions on the rear of BL. Values $\tau$ calculated according to (2) for some options are given in Table 1 with the most large value $\gamma E_{env}$; $w = 200$ W (see [7], Tables 5, 6).

The effective electric field $\gamma E_{env}$ creates a directed additive $\Delta p_i$ to momentum of an ion on the back edge of BL. Every ion runaway by the field force takes with him this added momentum and does not create recoil force $e\gamma E_{env}$ on the rear edge of BL. Full momentum $\Delta P$ transmitted per time unit to the surrounding air molecules due to recoil collisions of ions with them in the rear of BL is $\Delta P = \Delta p_i \cdot (n_i^* v_i) \cdot \pi r_{FB}^2$ where $n_i^* = n_i (1-\tau)$, and $v_i$ is mean velocity of ions toward the rear boundary of BL including created by electric field additive $\Delta v_i = \Delta p_i / m_i$. Calculated values $\tau$ would be of a few less at accounting for the additive in velocity $v_i$ than the values shown in Table 1.

Note that at the transition from the equilibrium radiation power $W$ to an observable radiation power of non-equilibrium plasma $w$, $W \to w = W/\alpha^2$, at $n_e \to x_e n_m (T_m) / \alpha = n_e^{(FB)}$ [7], where $n_m$ is density of all particles, the values of non-equilibrium parameter $\alpha$ can additionally decrease with growing $n_e^{(FB)}$ due to decrease of reabsorbtion by cold molecules at the given observed $w \sim (100-200)$ W, what can lead to a significant increase of $\gamma E_{env}$ in these tables (see also below Section 4).

Thus, in the most cases $\tau \ll 1$, what justifies the assumption of compensation of electrical force $|e|\gamma E_{env} N_i$ in the rear of BL acting on the ions due to recoil response of atmospheric air [7] ($N_i$ is ions content in BL).

Moreover, since BL can be regarded as a conducting body, due to equipotential volume BL in relation to the external field $E_{env}$ the electric field is distributed over the entire surface of BL (i.e. parallel force lines of an external electric field $E_{env}$ should be some more bent toward the center of BL), what provides a basis to characterize the leakage across the positive charged BL surface with coefficient $\tau \ll 1$. Greater flows of runaway electrons can lead to greater uncompensated charge of the ions. However, the extremely low coefficient of surface tension is not enough to keep the positive and negative charges from the expansion. Thus, the work function $A_e$ at typical nonequilibrium $n_e \sim 2 \times 10^{15} \text{cm}^{-3}$, $r_{FB} \sim 5$ cm is only 267 K [7]. The predominance of considerable volumetric positive charge of BL could cause an increase in the effective coefficient of surface tension, because it leads to an increase in electron work function $A_e \sim kT_e$. The excess of the positive charge $|e|\Delta N_i$ is determined by the coefficients of diffusion of electrons and ions $D_e$, $D_i$ and their mobilities $\mu_e$, $\mu_i$ [10]:

$$\Delta E \sim \frac{D_e - D_i}{|\mu_e| + |\mu_i|} \frac{1}{r_{FB}} \sim \frac{|e|\Delta N_i}{(2r_{FB})^2},$$

where $\Delta E$ is electric field of ambipolar diffusion, so at $T_e \sim 10\,000\,\mathrm{K}$, $T_i \sim 3000\,\mathrm{K}$, non-equilibrium $n_e \sim 2 \times 10^{15}\,\mathrm{cm}^{-3}$, $r_{FB} \sim 5\,\mathrm{cm}$, $\sigma_e^{(coll)} \sim 2 \cdot 10^{-16}\,\mathrm{cm}^2$, $\sigma_i^{(coll)} \sim 2.5 \times 10^{-15}\,\mathrm{cm}^2$, one has $D_e, |\mu_e| \gg D_i, |\mu_i|$; $|e|\Delta N_i \sim 4 \times 10^{-2}\,\mathrm{C}$ at $|e|N_i \sim 0.17\,\mathrm{C}$, with the need to take into account $\Delta E$ in the original equations (see [7], Eq.(5) and below Eq. (3)). Note in this seemingly paradoxical situation: despite the considerable volumetric positive charge of BL, it moves in the direction of positive atmospheric electric field, attracting the electronic part of BL with smaller negative charge.

The presence of an uncompensated positive charge of the BL might sometimes be connected with the observed effect of bouncing BL [1] with conditioned by atmospheric electric field and polarization attraction of the negative charge of BL to the Earth's surface, and then the elastic repulsion of one-signed positive charges due to forming vertical $z$-component of electric field.

Equating the specific energy for bulk ionization and specific energy of bulk radiation at mean photon energy $\bar{I}$ and the electron density $n_e \equiv n_{e(eq)}/\alpha$, where $n_{e(eq)}$ is concentration of electrons at temperatures $T_m$, $T_e$; $p$=1at, we obtain

$$n_e n_m(T_m) I_0 \int_0^\infty \sigma_i v_e f_M dv_e \simeq n_e n_i \kappa_{rec} \bar{I} \simeq n_e (n_e + \Delta n_i) \bar{I} \frac{\varepsilon \sigma T_e^4 S_{FB}}{V_{FB} n_{e(eq)}^2 \bar{I}} = w \sim (100 - 200)\,\mathrm{W},$$

where $I_0$ is ionization potential, $w$ is observable radiation power, $f_M$ is normalized Maxwellian electron velocity distribution, $\varepsilon$ is degree of blackness [11], $\Delta n_i$ is excess of positive ions, neglecting the small contribution of free-free transitions at high $T_e$,

$$\kappa_{rec} \simeq \frac{\varepsilon \sigma T_e^4 S_{FB}}{V_{FB} n_{e\,eq}^2 \bar{I}}$$

is recombination coefficient, $k_{rec} \to \mathrm{const}$ at small optical path, $n_{e(eq)} \equiv n_e(T_e, 1\,\mathrm{at})$, $S_{FB}$ and $V_{FB}$ are surface area and volume of BL. From these equations one can estimate for a given $T_e$ and observable $w$ characteristic densities $n_e$, $n_m$, temperature of the neutral molecules and ions $T_m$ and non-equilibrium degree $\alpha$ in BL.

By using assumed free parameters: the diameter BL $d_{FB}$ or the external electric field $E_{env}$ in the original equation (3) (see below Section 3) which relates the size and the field strength [7], the luminosity $w$, while adding stability conditions (see below Eqs. (5), (6), (10)) and balance equations we can determine the entire set of parameters that characterize BL. However, it remains unclear whether the second parameter, e.g., the radiation power $w$, is free.

We note here that equation (3) obtained before [7] as a consequence of the equality of accelerations in the joint motion of dipole charges $N_e$, $N_i$ in an external electric field $E_{env}$, can describe unstable regime, maintaining a distance between the positive and negative charges but with the further its stabilization in the presence of balance of ionization and recombination and the excess positive charge.

We can also hypothetically assume the presence of augmented stability of separated charges with partly decreased their leaks because of geometric reasons, due to generally non-symmetrically smeared on the spherical surface unlocked layers of the condensations of electrons and ions at both ends of the plasma dipole, which can form observable unlocked shells on the BL border.

Due to recharge collisions of ions with neutrals and dipole attraction of charges of different signs, BL moves as an integer. It can be assumed that the internal coupling of the charges and their geometry, also excess positive charge may play a role of an additional surface tension, leading to a diversity of BL forms. One can image here that the entire plasma of BL is in a special state of counter ionization/recombination directed microflows disturbing Maxwell distribution. It ought also to be noted, that electrostatic forces are comparable with the Coulomb forces of interaction charged particles on average distances between them ([7], Table 7) and it can be accounted for by polarization factor $\gamma$ (see below Section 6).

Distribution of electron and ion densities along the radius is not Boltzmann (while maintaining the background Maxwell velocity distributions with the temperatures $T_e$, $T_m$) and is determined by the balance of recombination and ionization in the directed microflows of charged particles at a given current range of distance $r$. Assuming electrons being prevented by uncompensated positive charge from the repulsion and running away, we can assume that at the boundaries of BL in the regions of the dipole concentrations of electrons and ions, the spatial distribution can be still close to the Boltzmann distribution.

Local decrease in the density of electrons $n_e(r)$ in front of the leading edge of BL is determined by the Boltzmann distribution $\exp(-U(r)/kT_e)$, where at $r \sim r_{FB}$ we have $U(r) = 2|e|\gamma E_{env} r$. Thus, distance $r \gtrsim r_{FB}$, at which electron density decreases in $\exp(\beta_e)$ times in the interval $\Delta r \cdot r$ at a given value $\beta_e \sim (2 \div 4)$ is determined by relations

$$\frac{d}{dr}\, 2|e|\gamma E_{env} r \cdot \frac{\Delta r}{kT_e} \cong \beta_c, \quad \Delta r \sim \frac{\beta_c kT_e}{2|e|\gamma E_{env}} \quad \text{at} \quad r \gtrsim r_{FB}.$$

Thus, at $\beta_c \sim 2$ and BL parameters according to (9a), (9b) already at a slight excess of $r > r_{FB}$ we get smearing on the front electron end of BL $\Delta r \sim 0.2$ cm.

Considerable smearing of the visible front BL can occur also by diffusion of metastable molecules.

Consider the balance of ionization and lateral diffusion for more easily ionized oxygen component $O_2$ $(I_{O_2} \cong 12.2\ eV)$. Let us $T_e = 10\ 000$ K, $T_m = 2000$ K, $r_{FB} = 5$ cm. One obtains then $n_{O_2} = 7.50 \times 10^{17}\ cm^{-3}$, and for the rate of lateral ambipolar diffusion

$$dN_{diff}/dt \cong 4\pi n_e D_a r_{FB} \cong 3.6 \times 10^2 n_e,$$

$$D_a \cong \frac{\lambda \overline{v_{rel}}}{3}(1 + \frac{T_e}{T_{env}}) = \frac{1}{3}\frac{1}{\sigma_{coll} n_L}\frac{2}{\sqrt{\pi}}\sqrt{\frac{kT_{env}}{m_{O_2}}} \cong 5.7\ cm^2/s, \quad \sigma_{coll} \sim 2.5 \times 10^{-15} cm^2, \quad n_L = 2.5 \times 10^{19} cm^{-3}.$$

The ionization rate is given by

$$\frac{dN_i}{dt} = V_{FB} n_{O_2} n_e \int_{I_0}^{\infty} \sigma_{ion}^{(O_2)}(\varepsilon) v_e f_M(\varepsilon) d\varepsilon \cong V_{FB} n_{O_2} n_e \sigma_{ion}^{(O_2)}(I)\sqrt{\frac{2}{m_e}}\frac{2}{\sqrt{\pi}}\frac{(I + kT_e)}{(kT_e)^{\frac{3}{2}}}\, e^{-\frac{I}{kT_e}} \cong 1.1 \times 10^4 n_e,$$

$$f_M = \sqrt{\frac{2}{\pi}}\frac{\varepsilon^{\frac{1}{2}}}{(kT_e)^{\frac{3}{2}}}e^{-\frac{\varepsilon}{kT_e}},$$

and there is accepted near threshold ionization cross section $\sigma_{ion}^{(O_2)}(I) \sim 10^{-18} cm^2$, $I = 15\ eV > I_{O_2}$.

Thus, there is $dN_i/dt \gg dN_{diff}/dt$, regardless of the concentration of electrons $n_e$. At decreasing electron temperature the ionization rate decreases rapidly, and below $T_e \lesssim 8000$ K (if we ignore the retaining attraction by not compensated positive charge) BL is dissipated. Presence of an excess positive charge leads to additional keeping of electrons and expansion of possibilities to apply dipole model of BL at lower temperatures up to $T_e \sim 4000$ K. The lifetime of stable BL is limited then only by the stochastic volatility of the atmospheric electric field.

About basic problem of the simple approximated account for the polarization effect in plasma in the previous works [7, 8] and in the subsequent statement of this work by means of introduction of the new factor $\gamma$ see Section 6.

## 3. Stability of BL

Input parameters for the characteristics of BL are atmospheric electric field $E_{env}$ and possibly the conditions under which the initial seed discharge occurs with certain parameters and dimensions.

It seems obvious, the electric dipole in vacuum at a constant electric field is completely unstable (charges either collapse, or run away). As is noted in [7], the real stability of the dipole gas plasma in the form of BL is provided by the fundamental role of compensation of the ion backward force $f = -|e|\gamma E_{env} N_i$ by atmospheric recoil and energy loss by radiation, providing constancy of the speed of BL in the most realistic case of thermodynamic non-equilibrium BL with $T_e \gg T_i = T_m$.

It would seem that dependence of intensity of atmospheric electric field on radius of BL can be received from a condition of equality of acceleration of free electrons under the influence of this field and acceleration of the whole BL under the influence of the force of an attraction of ions by electrons and recharge collisions ion/neutrals.

But at the same accelerations and velocities of ions and electrons, energy and momentum transmission from electrons to ions must be defined by their mass ratio.

Crucial for the balance of forces in the DDM BL is relation of forces within BL and the force $f_e = |e|\gamma E_{env} N_e$ of the atmospheric electric field acting on BL electrons, that determines kinetics and integrity of BL.

Keeping integrity of BL when driving at a constant speed implies equality the force of atmospheric electric field $f_e$ and the Coulomb attraction force electrons/ions $f_{inn} = e^2 N_e N_i/(2r_{BL})^2 = f_e$.

However, even with a small acceleration of all BL it must occur under force $(M_{BL}/M_e)f_e$ which is fantastically large. BL integrity leads to the resolution of the paradox.

Coulomb interaction between electrons and ions, as before, occurs with the force $f_e$, but running electrons are tightly "tied" to BL by non-electric braking frictional force by orders of magnitude higher than $f_e$, i.e. due to phenomenally very fast transfer of momentum obtained by electrons from external electric field to ions and molecules of BL (mainly due to inelastic collisions of electrons with ions and molecules of BL). Such transfer of the electron momentum to BL occurs partly at BL moving in air with a constant speed, making up resistance losses at BL movement.

Accounting for BL plasma polarizability is here and further in replacing $E_{env} \rightarrow \gamma E_{env}$ ( $\gamma \ll 1$ is polarizability factor).

Thus, from this equal accelerations of electrons mass $M_e$ and BL mass $M_{BL}$, the occurrence should follow of absolutely impossible fantastic value of tension of dipole electric field $E_{inn}$ corresponding to intensity of atmospheric electric field $E_{env}$

$$E_{inn} = f_e = \frac{|e|N_i}{(2r_{FB})^2} \rightarrow E_{inn} = \frac{M_{BL}}{M_e}\gamma E_{env} \gg f_e, \quad \text{hence} \quad \gamma E_{env} = \frac{\pi}{3}\frac{m_e}{m_{av}}|e|\frac{r_{FB}}{n_m}n_e n_i \ll E_{inn}, \qquad (3)$$

where $E_{inn}$ includes now huge non-electric friction force; $n_m$ is density of massive particles, $m_{av}$ is average mass of BL particle.

The cause of this paradox is also the unreasonable unjustified imaging BL-dipole as non-overlapping positive and negative charges as a dumbbell, but not as the being partly superimposed clouds of electrons and ions with the dynamic properties of the interchanging bulk electrons/BL-particles momentums.

Electric field of dipole attraction of electrons and ions of BL can be really very small, so the stability of BL as the single whole can be caused mainly by collision redistribution of electrons momentum and energy to all BL particles. Therefore one must replace $E_{inn} = M_{BL}\gamma E_{env}/M_e$ for $E_{inn} = \gamma E_{env}$ in the electron and ion dipole edges of BL, what removes this paradox (that is, it corresponds to the resulting summed force of accelerated or without acceleration traction of ions, correspondingly BL, by electrons, but the whole BL is accelerated by non electric friction forces). It occurs at acceleration almost fully in inelastic collisions of electrons with BL molecules with increasing $T_e$, enhance of the ionization-recombination processes and equalizing bulk accelerations and possibly velocities of clouds of ions and electrons, in the presence of the surface tension Coulomb forces of excess volumetric positive charge supporting geometrical form and stability of BL. It provides permanent integrity of BL and prevents dispersion BL at growing $\gamma E_{env}$, but with the possibility of destruction of BL at sharp large changes of $E_{env}$. This process can be operated at accelerating BL.

The observed integrity of BL can only be explained by the fact that at BL movement with the same effective acting forces of atmospheric field on the electrons $|e|E_{eff} \equiv |e|\gamma E_{env}$ and the dipole interaction electrons/ions (the least force), the coupling BL with electrons ("friction clutch", the most force) at enough small acceleration is ensured by the necessary to equalize the accelerations $M_{BL}/M_e$ times larger the BL pushing force by mainly inelastic collisions of electrons with molecules of BL.

Similar transfer of excess momentum of electrons to BL mass occurs at uniform motion of BL and is spent on compensation for the loss of energy in the air resistance to its motion. However friction force is very specific, and if there is no acceleration, the electron cloud within BL is immovable relative to it, and the friction force is equal to zero.

Equations (3) can be clearly and easily understood as the necessary (but not sufficient!) conditions of existing BL as an integer whole due to the force balance.

Thus enough high rate of collision transmission and redistribution of excess momentum from electrons to BL as a whole is one of conditions of stability of BL in acceleration modes.

The second basic stability condition can be indicated existence of minimum of dependence, for example, of BL sizes and potential energy on electron temperature $T_e$.

The relation (3) where we neglected a small ion excess $N_i \gtrsim N_e$ can be rewritten as

$$r_{FB} = \frac{3}{\pi}\frac{m_{av}}{m_e}\frac{\gamma E_{env}}{|e|}\frac{n_m}{n_e^2}\alpha^2 \equiv A\frac{W}{n_e^2}, \quad A \equiv \frac{3}{\pi}\frac{m_{av}}{m_e}\frac{\gamma E_{env}}{|e|}\frac{n_m}{w}, \qquad (3a)$$

where $n_m$ is density of neutral molecules and atoms in BL; $n_e = x_e n_m$; $\alpha \geq 1$; $w$ is observable thermodynamically non-equilibrium radiation power of BL adopted in options $w$ to be equal to $\sim (100 - 200)$ W [7]; $m_e$, $m_{av} \simeq m_i$ are correspondingly mass of electron and average mass of neutral in BL. Generally, stability of BL is caused by presence of two free parameters. We will consider as illustration simplified variant of the fixed free parameter $w = \text{const}$ and variable free parameter $T_e$ at $\gamma E_{env} \sim \text{const}$, $n_m(T_m) = \text{const}$, $A \sim \text{const}$. At large optical depths we obtain

$$W = 4\pi r_{FB}^2 \varepsilon \sigma T_e^4, \quad \varepsilon \sim \varepsilon(T_e,\ 10\ \text{cm})\sqrt{\frac{r_{FB}}{10\ \text{cm}}}, \qquad (4)$$

$\sigma$ is Stefan-Boltzmann constant, $\varepsilon$ is here the degree of blackness of the hemispherical volume [11], $\alpha^2 \equiv W/w = (n_e/n_e^{(FB)})^2$, $W$ is power of equilibrium radiation BL, $n_e \equiv n_e(T_e)$ is the equilibrium density of electrons, $n_e^{(FB)}$ is actual non-equilibrium electron density of BL. All abovementioned does not exclude possible accelerating movement of BL as a whole.

It is assumed at definition of the non-equilibrium parameter $\alpha$ that the balance of charged-particle densities $b_r n_e^2 = b_i n_e n_m$, where $b_r$ is coefficient of recombination, $b_i$ is ionization coefficient, $n_m$ is concentration of neutrals, i.e., $n_e = n_m b_i / b_r$, leads to a sharp decrease $n_e$ due to predominant compared to $b_r \sim \mathrm{const}$ sharp decrease $b_i$ with decreasing temperature $T_{FB} \equiv T_m < T_e$.

Note that the derivation of equation (3) neglects the very small gasdynamic resistance of the medium [7], which affects the BL as a whole if one images BL as a solid body.

With a fixed value of $E_{env}$ we obtain necessary derivative condition for stability of BL

$$\Theta \equiv \frac{dr_{FB}}{dT_e} = \frac{AW}{n_e^2} \frac{[\partial(\ln \varepsilon T_e^4)/\partial T_e - 2\partial(\ln n_e)/\partial T_e]}{1 - A\frac{W}{n_e^2}\frac{\partial(\ln \varepsilon r_{FB}^2)}{\partial r_{FB}}} = 0, \qquad (5)$$

and with obtained further (see below) according to (5) temperature $T_e \simeq 12\,500\,\mathrm{K}$

$$1 - A\frac{W}{n_e^2}\frac{\partial(\ln \varepsilon r_{FB}^2)}{\partial r_{FB}} < 0, \qquad (6)$$

$$\frac{\partial(\ln \varepsilon r_{FB}^2)}{\partial r_{FB}} \simeq \frac{\partial}{\partial r_{FB}}[\ln(r_{FB}^{5/2})] = \frac{5}{2r_{FB}}, \qquad (7)$$

when for large optical depths $\varepsilon(T_e, r_{FB}) \propto r_{FB}^{1/2}$. Thus, equation (5) reduces to the equation

$$\partial(\ln \varepsilon T_e^4)/\partial T_e - 2\partial(\ln n_e)/\partial T_e = 0. \qquad (8)$$

| Table 2. To solving the stability equation (5) | | | | | | | |
|---|---|---|---|---|---|---|---|
| $T_e, \mathrm{K}$ | 6000 | 7000 | 8000 | 9000 | 10 000 | 11 000 | 12 000 |
| $\lg \varepsilon T_e^4$ | 12.155 | 12.494 | 12.818 | 13.294 | 13.832 | 14.310 | 14.714 |
| $n_e, \mathrm{cm^{-3}} \times 10^{-15}$ | 1.27 | 4.39 | 17.2 | 60.9 | 171 | 392 | 762 |
| $\lg n_e^2 - 30$ | 0.208 | 1.284 | 2.472 | 3.570 | 4.466 | 5.186 | 5.764 |

| Table 3. To solving the stability equation (5) | | | | | | | |
|---|---|---|---|---|---|---|---|
| $T_e, \mathrm{K}$ | 6500 | 7500 | 8500 | 9500 | 10 500 | 11 500 | 12 500 |
| $x = \Delta \lg \varepsilon T_e^4$ | 0.336 | 0.323 | 0.377 | 0.538 | 0.478 | 0.404 | - |
| $y = \Delta \lg n_e^2$ | 1.076 | 1.188 | 1.098 | 0.896 | 0.720 | 0.578 | - |
| $y - x$ | 0.740 | 0.865 | 0.720 | 0.358 | 0.242 | 0.174 | ~ 0 |

As an example, consider the robust option of BL, as calculated with the following option parameters [7]:

$$T_e = 10\ 000\ K; \quad T_i = T_m = 1000\ K; \quad r_{FB} = 10\ \text{cm}; \quad \varepsilon = 6.8 \times 10^{-3}; \quad \alpha \approx 49.3; \quad \gamma E_{env} = 46.6\ \text{V/cm}; \quad (9a)$$

$$v_{FB} = 1.85 \times 10^{-2}\ \text{m/s}; \quad n_e^{(FB)} = n_e / \alpha = 3.47 \times 10^{15}\ \text{cm}^{-3}; \quad w = 200\ \text{W}. \quad (9b)$$

The results of numerical calculations are presented in Tables 2 and 3.

Comparison of the differences $\Delta \lg \varepsilon T_{FB}^4$ and $\Delta \lg n_e^2$ confirms presence of the solution of stability condition (5) with $n_e^{(FB)} = n_e / \alpha$ ($n_e = x_e n_m$ from [9]) at temperature $T_e \simeq 12\ 500\ K$, which should be considered a true for this option instead of arbitrarily chosen prior value $T_e = 10\ 000\ K$. With the given values $w$ and $E_{env}$, at temperature $T_e = 12\ 500\ K$, from the relation (3a) we can obtain corresponding to it new values $\varepsilon$ and $\alpha$ with moving BL in its parametric space of two free parameters $w$ and $E_{env}$ with defined in Section 6 polarizability factor $\gamma$.

The additional second condition for stability

$$\frac{d\Theta}{dT_e} = \frac{\partial\Theta}{\partial T_e} + \frac{\partial\Theta}{\partial r_{FB}} \frac{dr_{FB}}{dT_e} > 0 \qquad (10)$$

confirms stability of BL in the minimum point.

On the contrary, we can put the issue of stable value $r_{FB}$ at the fixed field tension $\gamma E_{env} = 46.6\ \text{V/cm}$ and a fixed temperature $T_e = 10\ 000\ K$ by solving the equation

$$\frac{dT_e}{dr_{FB}} = 1 - \frac{AW}{n_e^2} \frac{\partial[\ln(r_{FB}^{5/2})]}{\partial r} = 0 \qquad (11)$$

we have the result

$$r_{FB} = \frac{n_e^2}{10\pi A\varepsilon\sigma T_e^4} \simeq 4.2\ \text{cm} \qquad (12)$$

with increasing $r_{FB}$ up to $r_{FB} = 10\ \text{cm}$ at replacement of the fixed temperature by $T_e = 12\ 500\ K$. It means presence of the second branch of solution.

This result can be interpreted as two-dimensional parametric space $(T_e, \alpha)$, or $(T_e, w)$, or $(r_{FB}, w)$ of continuous series of stable ball lightning where $T_e$ and $w$ are related with trajectory of the stability condition (5).

When the electron density is much less than equilibrium value, reabsorbtion may be disregarded and instead of Eq. (4) one can use the approximation

$$\varepsilon \sim \varepsilon(T_e,\ 1\ \text{cm}) \cdot \frac{r_{FB}}{1\ \text{cm}}, \qquad (4a)$$

then instead of Eq. (7) will be

$$\frac{\partial(\ln \varepsilon r_{FB}^2)}{\partial r_{FB}} \simeq \frac{\partial(\ln r_{FB}^3)}{\partial r_{FB}} = \frac{3}{r_{FB}}. \qquad (7a)$$

However now the rate of increase $\varepsilon$ additionally grows with increasing temperature $T_e$, what will lead to increased values $x$ in Table 3 and consequently to lowering the temperature $T_e$ at the point of minimum $\partial r_{FB}/\partial T_e = 0$ appropriate to stable BL.

For the derivative of the total potential energy of ball lightning at, for instance, the constant free parameters $E_{env}$ and $\alpha$

$$U_t \sim 2|e|\gamma E_{env} r_{FB} \times \frac{4}{3}\pi r_{FB}^3 \cdot \frac{n_e}{\alpha} \equiv C r_{FB}^4 n_e; \quad C = \frac{8}{3}\frac{\pi}{\alpha}|e|\gamma E_{env}$$

we have:

$$\frac{dU_t}{dT_e} \sim 4C r_{FB}^3 n_e \frac{dr_{FB}}{dT_e} + C r_{FB}^4 \frac{dn_e}{dT_e}.$$

Comparing the two terms on the right side, we obtain

$$\frac{C r_{FB}^4\ dn_e/dT_e}{4C r_{FB}^3 n_e\ dr_{FB}/dT_e} \equiv \eta = \frac{r_{FB}}{4} \frac{dn_e/dr_{FB}}{n_e}. \qquad (13)$$

Hence, expressing $n_e$ in Eq. (5) through $r_{FB}$ with the same $\alpha$ we have $\eta = -1/8$. It means that if $dr_{FB}/dT_e \to 0$ also $dU_t/dT_e \to 0$, and condition $dr_{FB}/dT_e \to 0$ is equivalent to the condition of minimum potential energy. It should be noted, however, that the level of stability also depends on the depth of the potential well.

A similar procedure can be applied in determining the stability of BL with thermodynamic equilibrium plasmas for a model with one free parameter

$$n_m = n_L \frac{T_{env}}{T_e}, \quad n_e \ll n_m, \qquad (14)$$

that is at

$$r_{FB} = B \frac{T_e}{x_e^2}, \quad B \equiv \frac{3}{\pi} \frac{m_{av}}{m_e} \frac{\gamma E_{env}}{|e|} \frac{1}{n_L T_{env}}, \qquad (15)$$

where $x_e \equiv x_e(T_e)$ is molar fraction of the electrons [9], $n_L = 2.5 \times 10^{19}\,\mathrm{cm}^{-3}$. In this case at least at $T_e \lesssim 12000$ K,

$$\frac{dr_{FB}}{dT_e} = \frac{B}{x_e^2}\left(1 - 2T_e \frac{d\ln x_e}{dT_e}\right) \ll 0. \qquad (16)$$

Thus, previously considered BL with thermodynamic equilibrium plasma [7], in addition to that it has unrealistically huge radiation power, is unstable because the derivative does not pass through null. However, the existence of BL depends not only on compliance with expressions of the type (5), (7), (10), (12) or (16), but on the admissibility of arbitrary (in fitting to subjectively observed values $w$) choose the degree of non-equilibrium in the form of the parameter $\alpha$. Allowable non-equilibrium is ultimately determined by the balance of rate of recovery the non-equilibrium state of the plasma and the leakage rate. In this case

*Table 4. Effect of leakage of charged particles and possible reabsorbtion on the parameters of BL ( at accounting for the loss of power to the creation of the front jet, the speed should be doubled)*

| $r_{FB}$, cm | $T_e$, K $\times 10^{-3}$ | $T_m$, K $\times 10^{-3}$ | $\varepsilon \times 10^3$ | $x_e$ [9] $\times 10^3$ | $\alpha$ | $\beta$ | $n_e^{(FB)}$, cm$^{-3}$ $\times 10^{-15}$ | $E_{env}$, V/cm | $F_{FB}$, kgf | $v_{FB}$, m/s |
|---|---|---|---|---|---|---|---|---|---|---|
| 5 | 8.5 | 3 | 1.8 | 4.6 | 5.93 | 2.4 | 1.9 | 40.6 | 65.2 | 0.31 |
| 5 | 8.5 | 3 | 1.8 | 4.6 | 4.4* | 3.2 | 2.56 | 74.4 | 163 | 0.13 |
| 10 | 8.5 | 3 | 2.0 | 4.6 | 12.7 | 2.2 | 0.884 | 17.7 | 102 | 0.19 |
| 5 | 10 | 3 | 5.0 | 23 | 11.5 | 6.0 | 4.88 | 308 | 1280 | 0.016 |

* $\theta = 1.5$; $W/w = \alpha^3$; $w = 200$ W; $n^{(FB)} = n_e/\alpha$.

the main source of replenishment of electrons is ionization of relatively cold molecules with electrons under the influence of the internal field $E_{inn} \sim \gamma E_{env}$ and the loss of electrons due to recombination radiation.

Changing the external conditions for maintaining stable BL may lead in some cases to development of all sorts of turbulent instabilities that precede the destruction of BL with the appearance of the visible complex structure of internal inhomogeneities.

## 4. Effect of leakage of charged particles, metastable molecules and reabsorbtion

Leakage of electrons, ions and metastable excited molecules, also reabsorbtion lead to the need to reduce $W$, what one can account with replacement $\alpha^2 = W/w \to \alpha^{2\theta}$, $\theta > 1$, where, for example, for illustration one can take $\theta \sim 1.25$, i.e.

$\alpha = (W/w)^{0.4}$, $w \sim 200$ W, $n_e \to n_e^{(FB)} = n_e/\alpha$. Then the calculation according to equation (3) illustrated by Table 4 shows the strong increase $\gamma E_{env}$ at a relatively small increase $\theta$ (in Table 4 $\beta$ characterizes the share $\xi = 1 - e^{-\beta}$ of all BL molecules which are entrained by BL at its movement in air [7]).

With increasing $\theta$ negative value of the denominator in (11) tends at some $\theta \to \theta_0$ to zero

$$1 - \frac{AW^{1/\theta}}{n_e^2} \frac{\partial \ln(\varepsilon r_{FB}^2)}{\partial r} \to -0. \qquad (17)$$

Thus, with increasing leaks over certain critical value $\theta > \theta_0$ BL does not exist. Equating the specific power of radiation losses (as a major loss)

$$W_\varepsilon = \frac{r_{FB}}{r_0} \varepsilon(r_0) \sigma T_e^4 \frac{(n^{(FB)})^2}{n_e^2} \frac{S(r_{FB})}{V(r_{FB})} \equiv (n^{(FB)})^2 \Phi_\varepsilon, \qquad (18)$$

where $r_0$ is radius, which falls in the region of small optical densities with a linear dependence of the emissivity $\varepsilon$ on $r$, and the specific energy expended per unit time for ionization

$$W_i = n_e^{(FB)} \left[ n_m^{(O_2)} \int_{I_{O_2}}^{\infty} \sigma_i^{(O_2)} v E f(E) dE + n_m^{(N_2)} \int_{I_{N_2}}^{\infty} \sigma_i^{(N_2)} v E f(E) dE \right] \equiv n_e^{(FB)} \Phi_i, \qquad (19)$$

$$f(E) = \frac{2}{\sqrt{\pi}} \frac{E^{1/2} e^{\frac{E}{kT_e}}}{(kT_e)^{3/2}}, \qquad (20)$$

where $\sigma_i^{(O_2)}$, $\sigma_i^{(N_2)}$ are ionization cross sections of electrons with a temperature $T_e$ in collisions with air molecules with temperature $T_m$, we obtain

$$\alpha \equiv \frac{n_e}{n_e^{(FB)}} = \frac{n_e \Phi_\varepsilon}{\Phi_i}, \qquad (21)$$

and from comparison $W_\varepsilon$ with the equilibrium value $W_\varepsilon(T_e, n_e, r_{FB})$ we obtain the value $\theta$.

## 5. Summary on the main comparative features of DDM and BL

DDM well agrees qualitatively with the most of observations.

1. BL movement is generally directed, sometimes zigzag, in a wide range of speeds, and apparently is not determined by wind speed, but presumably by predominant influence of atmospheric electric field along its force lines whose direction (vertical, inclined or horizontal) depends on landscape and other local and general (position and the special features of the thunderstorm cloud) conditions with positive in the whole charge of the Earth.

2. DDM BL movement can only occur in the positive direction of the electric field (electron edge forward). It corresponds to sometimes observed vertical fall of BL to the Earth.

3. BL is relatively "cold" at a temperature $T_{FB} \sim (2000 - 4000)$ K, but sufficient hot at a radius of $r_{FB} \sim (5 - 10)$ cm for the observed burning holes in the window glass, with a relatively weak luminescence (100 - 200) W indicating thermodynamic non-equilibrium BL with electron temperatures $T_e \gg T_{FB}$.

4. The apparent passage of BL through intact window glass can be attributed to strong dipole polarization of DDM.

5. BL can occur without regard to the linear lightning.

6. Very strong local ambient atmospheric electrical fields $\gtrsim 10$ kV/cm (see below Tables 5 - 8) supporting existence of DDM BL can lead to lethal defeat of people and animals by electric current owing to facilitating breakdown at contact with BL.

7. Behind the DDM BL one could observe reactive recoil jet stream of cold or slightly heated atmospheric air due to the recoil force $f \sim |e| \gamma E_{env} N_i$ (see Section 6).

8. The lifetime of DDM BL can be determined by external factors, changes in the atmospheric electric field in time and space. However there are also internal intrinsic for BL factors related to instability that occurs at strong accelerating or decelerating movement of BL.

9. BL much diversity can be explained by the possible presence of two free parameters DDM, for example, radius and luminosity.

| Table 5. Constraint features of BL parameters | | | | |
|---|---|---|---|---|
| $T_m = 1000$ K, $T_e = 10\,000$ K, $(\gamma E_{env}) = 1$ kV/cm | | | $T_m = 3000$, $T_e = 10\,000$, $(\gamma E_{env}) = 1$ kV/cm | |
| $r$, cm | 10 | 100 | 1000 | 10 | 100 |
| $n_e^{(FB)}$, cm$^{-3} \times 10^{-15}$ | 16.0 | 5.1 | 1.6 | 9.2 | 2.9 |
| $\alpha$ | 10.6 | 33.7 | 107 | 6.2 | 19.6 |
| $E_{spec}$, J/m$^3$ | $5.15 \times 10^7$ | $1.63 \times 10^8$ | $5.12 \times 10^8$ | $2.9 \times 10^7$ | $9.2 \times 10^7$ |
| $\varepsilon$ | $6.8 \times 10^{-3}$ | $4.0 \times 10^{-2}$ | $\sim 2 \times 10^{-1}$ | $6.8 \times 10^{-3}$ | $4.0 \times 10^{-2}$ |
| $w$, kW | 4.3 | $2.5 \times 10^2$ | $1.2 \times 10^4$ | 12.5 | $7.4 \times 10^2$ |
| $w/S_{FB}$, W/cm$^2$ | 3.45 | 2.0 | 1.0 | 10 | 5.9 |
| Maximal TNT equivalent of BL | 0.052 kg | 0.164 t | 0.50 kt | 0.029 kg | 0.092 t |
| $E_{env} \gtrsim |e| \left[ n_e^{(FB)} \right]^{-2/3}$, $\frac{\text{kV}}{\text{cm}}$ | 9.1 | 4.3 | 2.0 | 6.3 | 2.23 |
| $\gamma$ | <0.11 | <0.23 | <0.51 | <0.16 | <0.45 |

10. There is characteristic blurring of BL and BL DDM back and forward edges of a few mm sizes.

11. For realizing DDM there are presumably required simple natural principles of self-organizing: (1) seed volume discharge with a high level of thermodynamic nonequilibrium $T_e \gg T_{FB}$. Most suitable for this it appears to create experimentally the seed microwave VHF discharge. The possibility of the spontaneous appearance of the seeding discharge with the local electrical breakdown in the places of sharply inhomogeneous local atmospheric electric field up to $E_{env} \gtrsim 10$ kV/cm at short distances is not also excluded; (2) strong and extended electric field driving the seed discharge to the stable translational movement.

12. The rapid movement of the observed big BL (may be accelerating?) is accompanied by a hissing sound, like a jet engine noise [2], perhaps, due to the recoil jet in the back of BL and possible joining the powerful local discharge.

13. BL in the region of strong atmospheric electric field may be a trigger of the accompanying local atmospheric discharge, highly increasing the destructive power of BL discharge, it even might be progenitor of the local discharge of linear lightning.

14. It should be noted that BL, strictly speaking, should be extended in an oval shape, characteristic for the dipole, with heterogeneous inner structure, and Fig. 1 is an idealization of the first approximation. Excess volumetric positive charge of BL keeping electrons from running away allows considerable observed variability in the form of BL throughout its movement along while maintaining its stability. However when my personal conversation the witness claimed that BL flying by him had not a specific shape and was highly luminous strongly pulsing object.

15. From Table 4 follows that at given $r_{FB}$ there is trend to significant decreasing BL velocity while increasing the temperature and energy density of BL (cf. also Tables 7 and 8).

16. Destructive power of explosion in DDM of BL is defined not only by BL energy content, but in not smaller degree by the rate of sudden falling of atmospheric electric field and BL power supply with resulting dipole collapse and the subsequent burst of quasineutral plasma.

17. Destruction of BL with its dispersion or the burst collapse can occur with the very rapid change of $E_{env}$ after which the slow damping action of collision processes have not time to occur.

_______________________

With the uniform translational movement of BL in an external electric field both the electrons and ions acquire energies, however at usually moderate speeds BL just a small part of the inside energy of BL falls to maintaining BL movement. Its most part is redistributed to the particles heating $kT_{FB}$ and $kT_e$ in the ion and electron collisions with molecules of BL and with each other and being lost in the form of radiation and power of forward recoil jet (see below). However, for implementing ionization and recombination processes inside BL there is need significant overlapping of ion and electron clouds, facilitated by sufficiently high temperatures $T_i = T_m = T_{FB}$ and $T_e$.

The assessments of parameters of the dipole dynamical model (DDM) BL along with similar estimates [7] show their qualitative agreement with the most part of the diverse phenomenological observations of BL.

Ball lightning is considered as a peculiar kind of the moving electric dipole with the balance of ionization and recombination in the middle of the dipole. Moving gas-plasma dipole is not directly an analogue of the static glow discharge confined in a finite space with virtual cathode and anode. In view of the chaotic thermal motion of electrons in the fireball, a balance of ionizations and recombinations is setting without the usual electric current between the virtual cathode and the anode (or rather weak one); ion and electron clouds at the ends of the dipole play the role of force shells of fireball together with its translational movement retaining BL from dispersion and collapsing. One can suppose that ionization by accelerated electrons prevails at the electron front of BL, recombination prevails in the back of BL, which accumulate the ions. However owing to integrity BL and a smoothing background of prevailing rarefied neutral gas component of BL, it is very difficult to image presence of sharp heterogeneity of BL. At the same time due to asymmetry of BL dipole, weak microflows of electrons and ions can arise between dipole ends which can appreciably disturb Maxwellian distribution.

## 6. Momentums, energies, polarization degree and the movement of BL

BL gets energy with the rate $|e|\gamma E_{env} N_e v_{FB}$ due to the movement and acceleration of electrons in the atmospheric electric field. Electrons are very fast randomized by the collisions with the neutrals and each other with an increase in the temperature $T_e$ which leads to the growth of the recombination rate and radiation losses with creating forward recoil jet. Radiation carries away only energy, but not momentum. At the same time the conditions of BL as the single whole and compensating the force of atmospheric electric field, which acts on the ions, indicate the additional energy supply by atmospheric electric field with the rate $|e|\gamma E_{env} N_i v_{FB}$ for the forming reactive jet in the rear end of BL. Atmospheric electric field ensures the summed power supply $2w$ of the losses of energy BL $|e|\gamma E_{env} N_e v_{FB}$ including radiation and forward jet power resulting in BL movement gasdynamic resistance, and also the same (at $N_i \approx N_e$) value of the energy supply by this external electric field for the creation of the backward reactive jet. An increase in the temperature $T_e$ and radiation leaks stops on reaching some value, which corresponds to stable moving BL.

If $m$ is mass of all electrons, $M$ is mass of entire BL, then according to the law of momentum conservation, i.e. with the equality of the momentum of translational electron movement (with velocity $v_1$) before, and BL (with velocity $v_2$) after the randomization of electrons, $mv_1 = Mv_2$ we will obtain for the energy of directed translational movement BL after the

$$Mv_2^2 \big/ 2 = \frac{m}{M}(mv_1^2/2) \ll mv_1^2/2.$$

Thus, practically entire initial energy of electrons $|e|\gamma E_{env} N_e s$ being obtained with the displacement BL up to the distance $s$ passes into the thermal energy of electron cloud with the temperature $T_e$. Atmospheric electric field supplies power finally to the reactive jet behind BL and the same power $|e|\gamma E_{env} N_e v_{FB}$ for the same time $t$ to the whole BL in the forward direction. Electrical energy received at movement BL is spent approximately for radiation with the formation of a brake recoil jet caused by it at forward electron edge of BL (i.e. brake force of uniformly moving BL at creating a forward recoil jet approximately equals the force of the back recoil jet and traction force and is $\sim F_{br}(v_{FB}) = 2w/v_{FB} = |e|\gamma E_{env} N_e$ with the doubled velocity $v_{FB}$).

The mechanism to stop increasing BL momentum at the movement of BL with constant velocity as a single whole can be explained by transmission of the forward directed momentum acquired by electron cloud to the resulting on recombination forward momentum of resulting neutral atoms and molecules with producing constant resistance to BL movement..

Such a way, the acquired forward momentum of electrons transforms to the forward momentum of neutrals. Then it remains to assume that the last ones transmit it to molecules of ambient cold air on the forward edge of BL, i.e. to the environment, which causes additional (besides the early taken gasdynamic resistance for spherical solid body) brake resistance to the movement of BL with the appearing paradoxical forward directed recoil jet of cold air. At movement of BL with constant speed, atmospheric electric field pulls electrons and tightly collisionally linked them positive ions together with the whole BL, with the force equal to the force of resistance to this movement. Thus, at BL movement with constant speed the energy of the outside source is spent approximately equally to energy supply of the backward jet, to radiation, to braking forward jet losses with the same jet momentum equal to the momentum of traction force. Collision processes imbalance can lead however to incomplete (self)compensating the momentum being transmitted to BL by external electric field and the same opposite momentum of gasdynamic resistance to the movement of BL, which can lead in one or another degree to the small acceleration of BL with the natural shortening of its lifetime due to arising instability.

Processes of ionization and recombination occur simultaneously on all the volume of BL. Note, that momentum directly transferred by electrons to ambient air molecules at movement of BL as a whole with keeping electrons by positive charge is very small (in the case of absent radiation) due to the relatively small collision cross section and small mass ratio $m_e/m_{mol}$. So, if the momentum transfer to heavy neutrals is realized only by electron collisions with them, then BL should accelerate by the force $f \sim |e|\gamma E_{env} N_e$ up to almost unlimited speed (up to BL destruction) without forming any forward jet and with strong amplifying backward jet.

Backward recoil jet can contain slight admixture of hot molecules and running away ions weakly decreasing movement velocity of BL. Moreover, we can assume that at a high intensity of the atmospheric electric field (but below the threshold of streamer breakdown) there is possible arising of the non-streamer atypical linear lightning in the form of BL with a train behind, forming an expanding channel of linear lightning.

There is also a significant attracting effect of excess positive charge of BL on retention of runaway electrons as equivalent of the surface tension.

The existence of this type of plasma formation not confined in any finite space is due to its forward movement with energy supplying by atmospheric electric field.

Behind the BL must be a reactive wake trace of surrounding air caused by the recoil momentum of BL. However, this recoil momentum appears to be strongly numerically overestimated, which requires further explanation and may be caused by incomplete polarization affecting of the atmospheric electric field force acting on the ions in the whole with corresponding decreasing of the resulting atmospheric electric field force acting on the electrons in the whole. This will lead to a corresponding decrease in the atmospheric electric field effective force acting on the electrons and an increase in the electric field necessary to support BL, with appropriate amendments to Eq. (3). We can assume that in this case it ought to enter a correction polarizability factor $\gamma$ to the field tension $E_{env}$ in (3) and replace in all other equations $E_{env} \rightarrow (\gamma E_{env})$, $\gamma \leq 1$ with the increase of the external atmospheric electric field at previous calculations on the $1/\gamma$ times. Key parameter in the value $\gamma$ is the ratio of $|e| E_{env} \equiv a_1$ to $a_1/a_2 \sim E_{env}/|e|\left[ n_e^{(FB)} \right]^{2/3} \lesssim 1$, then $\gamma \sim a_1/a_2$, and if $a_2/a_1 \lesssim 1$, then $\gamma \sim 1$. At $\gamma \gg 1$, BL should dissipate. From Table 7 [7] follows that $\gamma^{-1}$ can reach values $\gamma^{-1} > (10 \div 100)$. Thus, fireball plasma polarizability is very significant basic characteristic of BL. With increasing $r_{FB}$ and decreasing $n_e^{(FB)}$, $\gamma$ can increase up to 1 and somewhat more. All previous results are saved at a single condition of replacing in all the formulas and calculations $E_{env}$ by "effective" electric field ($\gamma E_{env}$) with a significant increase of the real atmospheric field $E_{env}$ up to $\gtrsim 10$ kV/cm scale (see Tables below). For large diameters, with some increase $n_e$ in the typical range $n_e \sim (10^{15}-10^{16})$ cm$^{-3}$, there might seem possible significant increase of $E_{env}$ to hundreds or thousands of with increasing energy density of BL by orders of magnitude to $J \lesssim 10^{11}-10^{12}$ J/m$^3$, but from Tables 5 and 7 there follow constrains with the much more low energy density bounds.

The potential energy of ball lightning $U \leq |e|\gamma E_{env} N_e s$ in atmospheric electric field, where $N_e$ is content of electrons in BL, $s$ is arbitrary running distance of BL in the electric field, is consistently converted into radiation energy. Thus, depending on $s$, the values of potential energy can be arbitrarily large with the partial ratio of internal potential energy and internal kinetic energy of ball lightning which is not related to the virial theorem. Ratio of the inner potential and kinetic energies of BL is defined by no virial theorem but is a function of BL energy loss rate including radiation. The virial theorem is inapplicable to BL since BL is only part of the large system with an extended atmospheric electric field, radiation, and the reactive recoil jet of air behind BL.

The mystery of ball lightning is untangled by the fact that it is unique, having no analogues plasma object which is moving in the atmosphere, can be highly polarized, highly nonequilibrium air plasma cluster, and is fueling with the internal energy by an infinite extended external electric field at BL movement.

Near the ion part of BL it should occur "boiling" recombination processes with decelerating microflows of fast $kT_e$-electrons from electron dipole end, correspondingly accelerating ions to rear BL end, and ionization process near electron dipole end, with

| | | | | |
|---|---|---|---|---|
| $r_{FB} = 5$ cm | $\varepsilon \sim 3.0 \times 10^{-3}$ | $\alpha = 18.7$ | $|e|\gamma E_{env} = 1.09 \times 10^{-11}$ dyn | $v_{FB} = 1.62$ m/s |
| $T_m = 3000$ K | $x_e = 8.3 \times 10^{-3}$ | $n_e/\alpha = 1.09 \times 10^{15}$ cm$^{-3}$ | $e^2(n_e/\alpha)^{2/3} = 2.43 \times 10^{-9}$ dyn | $w/S_{FB} = 0.32$ W/cm$^2$ |
| $T_e = 9000$ K | $n_m = 2.44 \times 10^{18}$ cm$^{-3}$ | $\gamma E_{env} = 6.8$ V/cm | $1/\gamma \gtrsim 2.2 \times 10^2$ | $E_{spec} = 1.2 \times 10^4$ J/m$^3$ |
| $w = 100$ W | $n_e = 2.0 \times 10^{16}$ cm$^{-3}$ | $F_{FB} = 6.2$ kgf | $E_{env} \gtrsim 1.5$ kV/cm | $E_{full} = 5.6$ J |
| | | $n_e^{FB} \equiv n_e/\alpha$; $E_{th} \sim pV/V = 1.01 \times 10^5$ J/m$^3$ | | |

small charge leakage microflows to the side edges of BL. These steady microflows can considerably disturb Maxwellian distribution.

Apparently, the MHD- and Si-cluster model versions of BL may also occur as only atypical short-lived low-energy plasma objects tightly associated with the linear lightning, borrowing its energy ([1], [2]).

Note that the BL is an indicator of the regions of strong electric fields and may be a trigger of local electric discharge.

In the experimental reconstruction of BL main efforts should be aimed to creating a large scale system of guide electrodes, reproducing the very strong atmospheric electric field capable of moving and supporting the bulk seed discharge.

However, the emergence of the seed discharge remains unclear. One can expect to associate it with the spontaneous discharge between local atmospheric opposite charged clouds, corona discharges on the earth facilities (masts, trees, constructions of various kind), with the presence of motes in the local gathering of atmospheric electric field lines, sometimes with the stroke of local linear lightning. In experiment, the air inside the bulk seed thermodynamically non-equilibrium discharge must be sufficiently rarefied (may be with additional heating from an external source) to provide translational motion of dipole in an external electric field by the forces of ion recoil in surrounding air.

From the foregoing it should be also a fundamental output:

The existence of ball lightning is caused by its translational movement with supplying energy from the atmospheric electric field.

BL can not be created experimentally as a *static* plasma formation.

To create BL one might create highly non-equilibrium seed discharge (may be non-equilibrium microwave VHF-discharge) and a sufficiently strong and extended direct constant electric field, "pulling out" nascent BL, what is itself quite a complex engineering task.

Along with the systematic calculation of the parameters of BL with different radii, atmospheric field $E_{env}$ and rates of thermodynamic non-equilibrium parameter $\alpha$, particular interest would represent calculating parameters of powerful BLs with large dimensions $\sim$ (1 - 3)m and more (if exist) at account for the refinements in this paper.

Possible task for the future is also to obtain experimental running BL in different gas mixtures, including air with water vapor (without and with droplets) and easily ionizing gas mixtures at various pressures and temperatures.

According to equation (3) the electric field in BL is proportional to radius $r_{FB}$ at a given non-equilibrium electron density $n_e$. It follows from this that if the typical density (according to above examples of BL Table 4) is $n_e \sim 2 \times 10^{15}$ cm$^{-3}$, specific electrostatic energy of BL with volume $V_{FB}$ is $E_{spec} \equiv E_{elst}/V_{FB} \sim |e| n_e \gamma E_{env} r_{FB} \propto n_e^3 r_{FB}^2$, that is with a hypothetical theoretically increase in the diameter of BL up to 1.5 m and corresponding increase in radiation power $w$, supposed only specific electrostatic energy of charge separation itself could make $E_{spec} \gg 10^{10}$ J/m$^3$ (as compared with those ones of examples given by [7] $E_{spec} > 10^8$ J/m$^3$). It would seem, it agrees with observations $E_{spec} \sim 3.4 \times 10^{10}$ J/m$^3$ [12], [13] and $E_{spec} \sim 3 \times 10^{12}$ J/m$^3$ [14]. However it is not confirmed by the presented concrete calculations with $E_{spec} \lesssim (10^8 - \text{few } 10^9)$ J/m$^3$.

BL mechanical momentum $M_{BL} v_{BL}$ is very small; the demolishing power of BL is due to explosive release of the prevailing very large electrostatic energy of charge clouds separation.

Considering not so huge increase of really observable luminosity $w$ at the great sizes of BL, it ought to assume that at keeping electron density in the range $n_e^{(FB)} \sim (10^{15} - 10^{16})$ cm$^{-3}$, the temperature must considerably decrease up to $T_e \sim 8000$K with strong reduction of radiation power $\propto \varepsilon T_e^4 r_{FB}^2$ and them corresponding non-equilibrium parameter $\alpha > 1$, $n_e^{(FB)} = n_e/\alpha$. It ought also to account for the very strong reabsorbtion of radiation at BL sizes $\sim$100 cm and more, depending on the optical depth.

If $W_0$ is equilibrium power of the radiation corresponding to temperature $T_e$ in absence of reabsorbtion, and $W_r$ is power of radiation at presence of reabsorbtion, at replacement $n_e(T_e)$ on $n_e(T_e)/\alpha$ it is possible to receive the relation accounting reabsorbtion (at distribution of higher excited molecular states according to temperature $T_e$), in a kind

$$w = \frac{W_0}{\alpha^2}\left(\frac{W_r}{W_0}\right) = \frac{W_r}{\alpha^2}.$$

Values $W_0$ are determined by linear extrapolation of available data according to the "grow curve" from the small depths to the large optical thicknesses. It is shown above that BL is stable only in the case of strong thermodynamic non-equilibrium $T_e \gg T_i = T_{FB}$ and electron densities much smaller than at thermodynamic equilibrium, also maybe with additively diminished low luminosities due to the reabsorbtion, which at large optical depths is proportional to the quadratic root of the depth.

Under these assumptions it appears plausible that in dependence on the sizes of BL and increasing electron density in the range of $n_{ev} \sim (10^{15} - 10^{16})$ cm$^{-3}$ and observable energy density might take up to $E_{spec} \sim 10^{12}$ J/m$^3$ because BL could also be accompanied with a joining local lightning discharge[5]. Hypothetically it is not excluded also existence of two BL temperature branches: one with small $\alpha > 1$, and the second high temperature one with large $\alpha \gg 1$. Discovery of two branches for as small and large BL sizes might become very intriguing result characterizing the diversity of BL forms.

These considerations lay a way to further calculating toy variants characterizing large size BLs, including rather cumbersome calculations of stability conditions.

At the same time the calculations within the framework of estimations of the proposed approximate theory according to

$$\gamma E_{env} \simeq \frac{\pi}{3}\frac{m_e}{m_{av}}|e|r_{FB}\frac{n_e^2}{\alpha^2 n_m}; \quad \alpha^2 = W/w; \quad E_{env} = |e|\left[n_e^{(FB)}\right]^{2/3}; \quad n_e^{(FB)} \equiv n_e/\alpha; \quad (3b)$$

$$E_{spec} \simeq |e|\gamma E_{env}\cdot 2r_{FB}\cdot n_e/\alpha, \quad (3c)$$

show that, for example, at constraint with $\gamma E_{env} = 1$ kV/cm and $E_{env}$ constrained by the tension of streamer breakdown $E_{env} \sim (1 \div 10)$ kV/cm (at very small distances breakdown tension reaches up to $\sim 30$ kV/cm) and $r_{FB} = 100$ cm, we obtain

$$(\frac{n_e}{\alpha}) \lesssim 1.9 \times 10^7 \sqrt{n_m/r_{FB}} \sim 5.1 \times 10^{15} \text{ cm}^{-3}; \quad E_{spec} \lesssim 6.0 \times 10^{-2}\sqrt{n_m r_{FB}} \sim 1.6 \times 10^9 \text{ J/m}^3.$$

Values $\alpha^2 = W/w$ and $E_{spec}$ are constrained as by a maximum tension of atmospheric electric field, by observations of BL radiation power $w$ and by BL's stability conditions. The assessments presented in Table 5 show the indicative order of maximum energy density of large scale BL as $E_{spec} \lesssim (10^8 \div 10^9)$ J/m$^3$. The speed of movement BL is defined by relation

$$v_{FB} = 2w/f, \quad f = \gamma|e|E_{env}N_e$$

and can be the smallest, less than $(1 \div 10)$ cm/s. Examples of quickly moving BL's with the small sizes, low $\gamma$ and the small energy content, corresponding to low temperature $T_e \lesssim 8000$ K, are given in Tables 5 and 6 of the work [8] and some examples are presented below in Tables 6 and 8. For the speed there exists a simple ratio

$$v_{FB} = \frac{3}{\pi}\cdot\frac{2\varepsilon\sigma T_e^4}{\alpha^2 e^2 r_{FB}^2}\cdot\frac{m_{av}}{m_e}\cdot\frac{n_m}{\left[n_e^{(FB)}\right]^3}$$

(with account for energy loss to create the front recoil jet).

This seeming paradox of small velocities of large BL could be solved with that in the case of small BL with seeming constant observed speed of movement there are relatively large radiation losses and it was possible to neglect braking inertial forces, but in the case of large BL at relatively less radiation losses there is increased role of acceleration (which however was neglected) caused by the inertial term. All BL can travel with the acceleration in accordance with the simple balance law for BL energy $\mathscr{E}$:

$$\frac{d\mathscr{E}}{dt} = |e|\gamma E_{env}N_e(t)v_{FB} - w(t) - W_{gd}(t) - W_{dl} - W_{th}(t) - \frac{d}{dt}\frac{M_{FB}(t)\cdot v_{FB}^2}{2} = 0, \quad (22)$$

$$v_{FB} = \int_0^t a_{acc}(t)dt; \quad F = |e|\gamma E_{env}N_e; \quad N_e = n_e^{(FB)}V_{FB}, \quad s = \int_0^t v_{FB}dt, \quad (23)$$

---

[5]Comparison: for example, the specific heat of combustion of gasoline equals $\sim 3.1 \times 10^{10}$ J/m$^3$. Full electrostatic energy of charge separation is $E_{full} = (\gamma E_{env})|e|N_e\cdot(2r_{FB})$. Then at typical $n_e \sim (10^{15} - 10^{16})$ cm$^{-3}$, $r_{FB} \gtrsim (5 - 10)$ cm, $I_0 = 15$ eV : $E_{spec} \sim (\gamma E_{env})|e|n_e\cdot(2r_{FB}) \gg n_e I_0$.

where $N_e = n_e^{(FB)} V_{FB}$; $a_{acc}$ is BL acceleration; $F$ is electrostatic force of atmospheric electric field; $s$ is distance running by BL; $w(t)$ is power of radiation losses; $W_{gd}(t)$ is power spent on gasdynamic resistance of BL including power $w$ of the forward jet, $W_d$ is the rate of diffusion energy losses, $W_{th}(t)$ is power of thermal losses, the last term of Eq. (22) accounts for losses of overcoming the inertial forces. The set stationary mode of translational movement of BL corresponds to $d\mathcal{E}/dt = dv_{FB}/dt = 0$.

BL parameters are defined by accounting for Eq. (3), balance (or rate of processes) relations and stability condition. Difficultly achievable problem thus is the finding of dependences of base parameters of BL $r_{FB}(t)$, $T_e(t)$, $\alpha(t)$, $T_{FB}(t)$, $n_e^{(FB)}(t)$, $w(t)$, $W_{gd}(t)$, $W_{th}(t)$ on time, corresponding to the changing energy content $\sim |e| \gamma E_{env} N_e(t) \cdot 2r_{FB}(t)$ with the subsequent solution of the differential equation (22) for $v_{FB}(t)$, where as the first approach it is appearing reasonable to use the perturbation theory with the initial values corresponding to speed $v_{FB}$ at $d\mathcal{E}(t)/dt = 0$, $dv_{FB}(t)/dt = 0$. Then the asymptotically divergent solutions are possible with the explosion or the damping BL. Thus it appears to be usual presence of greater or smaller acceleration of BL due to Eq. (22) with $d\mathcal{E}/dt = 0$, $dv_{FB}(t)/dt \neq 0$ and possibility of arising kinematical instability. All BLs can move with acceleration, at least on the initial and probably on the final stage with occurrence of instabilities. Thus, acceleration can be signature and the possible internal cause, besides changes of an atmospheric environment and changes of $E_{env}$, defining BL lifetime (observed lifetime up to tens seconds). It is necessary to conclude that BL parameters

*Table 7. Limited energies of a fireball at arbitrary way accepted highest value of $w/S_{FB} = 1$ W/cm$^2$ ; $T_m = 2000$ K*

| $r_{FB}$, cm | $T_e$, K | $n_e^{(FB)}$, $\times 10^{-15}$ cm$^{-3}$ | $x_e$ | $\varepsilon$ $\times 10^2$ | $E_{env}$, kV/cm | $E_{spec}$, J/m$^3$ | $E_{full}$, TNT equiv. | $\alpha$ | $1/\gamma$ |
|---|---|---|---|---|---|---|---|---|---|
| 10 | | 1.4 | | 0.16 | >1.8 | $5.8 \times 10^4$ | 0.058 g | 6.10 | 118 |
| 100 | 8000 | 0.49 | $2.35 \times 10^{-3}$ | 1.3 | >0.90 | $2.4 \times 10^5$ | 0.241 kg | 15.2 | 57.6 |
| 1000 | | 0.18 | | ~10 | >0.46 | $1.4 \times 10^6$ | 1.41 t | 48.2 | 18.7 |
| 10 | | 7.0 | | 2.5 | >5.3 | $8.5 \times 10^6$ | 8.54 g | 54.1 | 13.7 |
| 100 | 12 000 | 3.5 | $1.04 \times 10^{-1}$ | 10 | > 3.33 | $1.1 \times 10^8$ | 0.111 t | 108 | 3.10 |
| 1000 | | 1.76 | | ~40 | >5.4 | $1.3 \times 10^9$ | 1.31 kt | 217 | $\lesssim 1.0$ |

*Table 8. Parameters of low energy Ball Lightning at $T_m = 2000$ K, $n_m = 3.66 \times 10^{18}$ cm$^{-3}$*

| | $r_{FB} = 5$ cm | | | $r_{FB} = 10$ cm | | |
|---|---|---|---|---|---|---|
| $w$, W | 50 | 100 | 200 | 50 | 100 | 200 |
| $w/S_{FB}$, W/cm$^2$ | 0.16 | 0.32 | 0.64 | 0.04 | 0.08 | 0.16 |
| $T_e$, K$\times 10^{-3}$ | 6 | 8 | 10 | 6 | 8 | 10 |
| $\varepsilon \times 10^3$ | 1.0 | 1.4 | 6.0 | 2.0 | 3.0 | 13 |
| $\alpha$ | 5.2 | 6.0 | 16.3 | 19 | 30 | 68 |
| $n_e^{(FB)} \times 10^{-15}$, cm$^{-3}$ | 0.33 | 1.4 | 5.2 | 0.091 | 0.29 | 1.25 |
| $E_{env}$, kV/cm | 0.69 | 1.8 | 4.32 | 0.27 | 0.63 | 5.5 |
| $\gamma E_{env}$, V/cm | 0.42 | 7.56 | 41.7 | 0.064 | 0.66 | 12.2 |
| $E_{spec}$, J/m$^3$ | $6.7 \times 10^4$ | $5.24 \times 10^6$ | $1.04 \times 10^9$ | $5.3 \times 10^3$ | $1.84 \times 10^5$ | $1.46 \times 10^7$ |
| $E_{full}$, J | $3.5 \times 10^1$ | $2.74 \times 10^3$ | $5.43 \times 10^5$ | $2.2 \times 10^1$ | $7.7 \times 10^2$ | $6.1 \times 10^4$ |
| $F$, kgf | 0.12 | 9.2 | 181 | 0.04 | 1.3 | 102 |
| $v_{FB}$, m/s | 85.4 | 2.20 | 0.22 | 128 | 15.6 | 0.39 |

**\*Note**. *Unlike the previous value for the degree of blackness $\varepsilon$ it was approximately applied here twice the value $\varepsilon$ of hemispherical degree of blackness. Let's notice that as $\varepsilon$ depends on the geometrical form of a radiating body, in case of a spherical body blackness degree accepts intermediate value between $\varepsilon$ of the middle of hemisphere base and doubled this value. When accounting for the fact that the radiation and the forward recoil jet takes the summed power $2w$, the speed $v_{FB}$ of BL should be doubled. $E_{full}$ contains here only electrostatic energy.*

$n_e^{(FB)}(t)$, $r_{FB}(t)$, $w(t)$, $F(t)$ etc. can be unequivocally found with consecutive use Eq. (3), calculation of balance/rate of

processes relations of redistribution of supplied electric energy between terms of (Eq. (22)), ionization balance and stability conditions of BL at selected values of one or, more probably two, free parameters, e.g. $E_{env}$ and initial value of luminosity $w$. However it is extremely difficult to find analytical dependences of these interconnected parameters on time, therefore it appears that the most prospective advancing way is to calculate diverse reasonable toy variants with some given initial values which are close to the $dv_{FB}/dt \sim 0$ of the asymptotical limit (if exist), then to compare results with observations.

It is difficult to image the existence of limiting modes of BL similar to the presented in Table 5, because of their seeming very large values of the force $F$ of the electrostatic dipole separation of charges $F = |e| \gamma E_{env} N_e$ and the essentially growing role of the inertial forces, which means necessity of account for above mentioned polarizability correction factor $0.01 \lesssim \gamma \lesssim 1$ in Eq. (22) and Table 5 which relates with very large value $E_{env} \gtrsim 10 \ \text{kV/cm}$ giving naturalness to extremely large $F$'s.

From Table 6 it follows that lethality of a fireball is caused not most by BL as that, but by the atmospheric electric field $E_{env}$ of which indicator it is.

In addition to Tables 5 and 6 there is presented Table 7 of as much as possible parameters and achievable energy densities in DDM.

Using the basic formula DDM Eq. (3b) and polarization condition

$$\gamma E_{env} < E_{env} \sim |e| \left[ n_e^{(FB)} \right]^{2/3} \qquad (24)$$

we obtain Eq. (25) :

$$E_{spec} < \frac{2\pi}{3} \frac{m_e}{m_{av}} e^2 r_{FB}^2 \left( x_e \sqrt{\frac{w/S_{FB}}{\varepsilon \sigma T_e^4}} \right)^3 \left( n_L \frac{T_{env}}{T_m} \right)^2 . \qquad (25)$$

Taking the reasonable limiting value

$$w / S_{FB} < (w/S_{FB})_{\max} \sim 1 \ \frac{\text{W}}{\text{cm}^2} ,$$

we obtain the parameters and higher limits of BL energies, presented in Table 7.

At presence of very big BL's both with small, and with large specific energies, the lifetime of very big BL's with the large radiation power can be limited by a fast exhaustion of the environment atmospheric electricity stock.

## 7. Temperature $T_m$ of gas constituent atoms and molecules of BL

Without going into a detailed description of the processes of heat exchange of BL it is still possible to obtain for them some general qualitative notion, considering the balance equation of heat flow supplied to atoms and molecules by electrons with temperature $T_e$ and of energy losses in collisions of BL atoms and molecules at temperature $T_m$ with the molecules of environment air at BL borders. One can describe the energy supplied to BL atoms and molecules by electrons in all BL volume per unit time as:

$$f_1 = n_m \sigma_e v_e n_e^{(FB)} (kT_m - kT_{env}) V_{FB} ,$$

and the lost energy due to collisions of BL molecules with atmospheric air on BL borders as

$$f_2 = n_m \sigma_m v_m n_L (kT_m - kT_{env}) S_{FB} \lambda_m ,$$

where $\lambda_m = 1/\sigma_m n_m$ is mean free path of BL hot atoms and molecules in the outer layer of air surrounding BL. Equating these flows of energy, after simple transformation we get the phenomenological expression for temperature $T_m$ of atoms and molecules within BL:

$$T_m = \frac{T_{env} + \dfrac{r_{FB}}{3} T_e n_e^{(FB)} \sigma_e \dfrac{v_e}{v_m}}{1 + \dfrac{r_{FB}}{3} n_e^{(FB)} \sigma_e \dfrac{v_e}{v_m}} ,$$

$$v_e = \sqrt{2kT_e/m_e}; \ \ v_m = \sqrt{2kT_m/m_{av}}; \ \ n_e^{(FB)} = x_e n_m / \alpha : \ \ n_m = 2.5 \times 10^{19} \cdot T_{env}/T_m \ \ ,$$

where $m_{av}N_m$ is average particle mass of atoms and molecules in BL, $\sigma_e$ is some effective total cross section of energy transfer of electrons to BL molecules, including the transfer of energy from the electron impact excitation of vibrational levels and through exciting electron levels. According to literary data, this cross section should not be great.

By selecting values $\sigma_e$ for options $r_{FB} = 10$ cm, $T_e = 8000$ K, $T_m = 2000$ K and $T_e = 12\,000$ K (see Table 7), in the first case the received solution for $T_m = 2000$ K is $\sigma_e \sim 1.32 \times 10^{-19}$ cm$^2$. Using the same cross section $\sigma_e$ in the second variant $T_e = 12\,000$ K we obtain the temperature $T_m \sim 4278$ K with corresponding final parameters:

$$T_e = 12\,000 \text{ K}; \ T_m \sim 4278 \text{ K}; \ r_{FB} = 10 \text{ cm}; \ \gamma E_{env} = 212 \text{ V/cm};$$

$$1/\gamma \simeq 15.1; \ E_{env} = 3.2 \text{ kV/cm}; \ E_{spec} = 2.32 \times 10^6 \text{ J/m}^3; \ m_{av} = 4.11 \times 10^{-23} \text{g}.$$

It should be noted the credibility of obtained value $\sigma_e$, it complies with usual temperatures $T_e$, $T_m$ of low pressure (and low density) gas discharges and temperature rise $T_m$ at increasing $T_e$.

## 8. Polarization oscillations of BL

Let us consider, for instance, the assumed mode of the polarization perturbation with the oscillations of polarization factor $\gamma$ and constant $E_{env}$ with $N_e(r) = N_i(r) \simeq n_e^{(FB)}V_{FB}$ using simplified momentum equation for the perturbation $\Delta r_{FB}$ based on Eq. 3:

$$\frac{d^2(\Delta r_{FB})}{dt^2} \sim -\frac{1}{M_{BL}}\frac{dF_{FB}}{dr} \cdot \Delta r_{FB} = -\frac{\Delta r_{FB}}{M_{BL}}\frac{d}{dr}(|e|\gamma E_{env}N_e); \quad \gamma = \frac{\pi}{3}\frac{m_e}{m_{av}}\frac{r_{FB}}{n_m}\frac{E_{env}^2}{e^2}; \quad E_{env} \geq |e|\left[n_e^{(FB)}\right]^{2/3} = \text{const.} \quad (26)$$

<table>
<tr><td colspan="6" align="center">Table 9. Polarization frequencies for BL variants of Table 1 ($w = 200$ W)</td></tr>
<tr><td>$r$, cm</td><td>5</td><td>5</td><td>10</td><td>10</td><td>10</td></tr>
<tr><td>$E_{env}$, kV/cm</td><td>1.37</td><td>2.98</td><td>3.28</td><td>4.56</td><td>2.18</td></tr>
<tr><td>$T_e$, K</td><td>8500</td><td>10 000</td><td>10 000</td><td>12 000</td><td>12 000</td></tr>
<tr><td>$T_m$, K</td><td>3000</td><td>3000</td><td>1000</td><td>1000</td><td>3000</td></tr>
<tr><td>$n_e^{(FB)} \times 10^{-15}$, cm$^{-3}$</td><td>0.93</td><td>2.97</td><td>3.47</td><td>5.61</td><td>1.87</td></tr>
<tr><td>$w/S_{FB}$, W/cm$^2$</td><td>0.74</td><td>0.74</td><td>0.16</td><td>0.16</td><td>0.16</td></tr>
<tr><td>$E_{full}$, J</td><td>7.55</td><td>$1.30 \times 10^2$</td><td>$2.17 \times 10^3$</td><td>$9.18 \times 10^3$</td><td>$9.68 \times 10^3$</td></tr>
<tr><td>$\nu$, kHz</td><td>2.68</td><td>15.3</td><td>6.32</td><td>13.5</td><td>7.61</td></tr>
</table>

Here $M_{BL}$ is BL mass. The minus sign in (26) is chosen on the basis of assumed condition of stability BL at fixed $E_{env}$, that is the presence counter force of the equal magnitude which arises at an imaged arbitrary small shift $\Delta r_{FB}$. One must use the expression (26) with the minus sign, because the force of attraction ions/electrons in (26) equals $-|e|\gamma E_{env}N_i$.

In this case, we can obtain for the frequency of polarization BL oscillations assuming constancy of $M_{BL}$:

$$\nu_{osc} \geq \left[\frac{7}{12\pi}\frac{m_e}{m_{av}^2}\frac{n_e^{(FB)}}{n_m^2}\frac{E_{env}^3}{|e|}\right]^{1/2}, \qquad (27)$$

where $M_{BL} = \text{const}$ is BL mass; BL oscillates as a single whole, including neutral particles and charged particles; $n_e^{(FB)} = \text{const}$, $n_m(r_{FB})V_{FB}(r_{FB}) = \text{const}$ (modified volumetric Langmuir oscillations). In the case of parameters of typical BL given in Table 6 we obtain naturally expected high frequency $\nu_{osc} \sim 3.4$ kHz in the region of the acoustic frequencies almost above an audibility threshold $(4 - 7)$ kHz (the thin sound of the high key or nothing above the audibility threshold).

*Table 10. Polarization frequencies for BL variants of Table 7 ( $w/S_{FB} = 1$ W/cm$^2$ )*

| $r$, cm | 10 | 100 | 1000 | 10 | 100 |
|---|---|---|---|---|---|
| $E_{env}$, kV/cm | 1.8 | 0.9 | 0.46 | 5.3 | 3.33 |
| $T_e$, K | 8000 | 8000 | 8000 | 12 000 | 12 000 |
| $T_m$, K | 2000 | 2000 | 2000 | 2000 | 2000 |
| $n_e^{(FB)} \times 10^{-15}$, cm$^{-3}$ | 1.4 | 0.49 | 0.18 | 7.0 | 3.5 |
| $W$, kW | 1.26 | $1.26 \times 10^2$ | $1.26 \times 10^4$ | 1.26 | $1.26 \times 10^2$ |
| $E_{full}$, J | $2.19 \times 10^2$ | $9.05 \times 10^5$ | $5.28 \times 10^9$ | $3.20 \times 10^4$ | $4.15 \times 10^8$ |
| $\nu$, kHz | 3.27 | 0.69 | 0.156 | 37.1 | 15.6 |

It is necessary to note however, that this calculation has more likely illustrative value as the parameters used in calculation depend on the rate of final setting diverse transient processes (lifetimes) with their proceeding in time. So, for example, some accompanying higher order effects were not accounted for: oscillations of temperatures $T_e$, $T_m$ and oscillations of radiating ability, there was any way supposed constancy of $M_{BL}$ etc.

Note also the possibility of sounds produced by possible other modes and by reactive backward jet.

Relations (26) and (27) are obtained for internal fluctuations of BL irrespectively from the translational movement BL on a trajectory. The oscillatory instability can be expected which limits BL lifetime as a result of the bifurcation of two expected complex conjugate solutions for the frequency of translational vibrations of BL with increment instability of BL velocity.

## 9. On the estimation of Ball Lightning lifetimes

Exact calculation of BL parameter dependences on time represents extremely a challenge as demands simultaneous calculation of a set of various processes, including, along with diverse macroscopic processes, also probabilities of many collision processes of microscopic scale (compare Section 7) and a choice of initial conditions and is complex problem. However it is possible to make general idea about an order of values for rates of change of parameters of BL, first of all about its lifetimes.

Leaving in BL energy balance equation (22) main terms of inflow rate of atmospheric electric field energy and the rate of energy outflow for the account of radiation losses, doing linearization with small initial perturbation of speed $\Delta v_{FB}$ at replacing $v_{FB} \rightarrow v_{FB} + \Delta v_{FB}$, neglecting indirect (inexplicit) dependence of mass and other parameters BL on time/speed and $r_{FB}$ of the same order of value, we obtain an equation for determination of dependence $\Delta v_{FB}$ $t$ in a kind

$$\frac{d \ln|\Delta v_{FB}|}{dt} = \frac{F - \frac{d}{dt} M_{FB} v_{FB}}{M_{FB} v_{FB}} = \frac{-\frac{dM_{FB}}{dt}}{M_{FB}}, \qquad (28)$$

from which for BL lifetime (or time of perturbation damping) $t_{life}$ defined as reverse value of increment/decrement, we obtain the trivial result of the type

$$t_{life} \gtrsim \frac{M_{FB}}{|dM_{FB}/dt|} = \frac{1}{\xi} \simeq \frac{T_m}{|dT_m/dt|} \sim \frac{T_e}{|dT_e/dt|} \sim \frac{v_{FB}}{|dv_{FB}/dt|}, \qquad (29)$$

where $\xi$ is relative increase/decrease of BL mass $M_{FB}$ in a time unit.

At initial perturbation $\Delta v_{FB}(0)$, the further growth (or damping) $\Delta v_{FB}(t)$ to which there corresponds proportional change $M_{FB}(t)$, is setting as

$$\Delta v_{FB}(t) \sim \Delta v_{FB}(0) \exp[\beta(t)t]; \quad \beta(t) + t\frac{d\beta(t)}{dt} \rightarrow \beta \sim \frac{dM_{FB}(t)/dt}{M_{FB}(t)}; \quad t_{life} \sim \frac{1}{\beta(t)} \sim \frac{M_{FB}(t)}{M_{FB}(t)\xi} \sim \text{const}. \quad (30)$$

As it is expected according to Section 7, $T_m$ ranges in narrow limits $\sim (2000 - 4000)$ K; if $dT_m/dt \lesssim (100 - 200)$ K/s, then $t_{life} \gtrsim (10 - 20)$ s which well agrees with observations. But at typical BL speed of the order of some m/s, in this time BL can fly many dozens meters, i.e. BL lifetime is determined by competition of both kinematic instability, and change of atmospheric electric field $E_{env}$ in space (and, probably, time).

Thus, relative fluctuations of parameters BL, for example, speed $v_{FB}$ under the influence of a wind, a component of all some percents, can lead to destruction of BL during an order of dozens seconds.

Depending on the sign $dM_{FB}/dt$, small perturbation is either damping (at $dM_{FB}/dt > 0$), or (at $dM_{FB}/dt < 0$) exponentially increases with the subsequent destruction BL. It does not contradict to that at large negative $\Delta v_{BL} < 0$, BL either dissipates at gradual decreasing of energy supply, or can collapse with explosion at sharp reduction of speed.

These preliminary considerations seem reasonable though are not valid as the strict proof.

If to assume that the part of power $d\mathcal{E}_{FB}/dt$ spent for acceleration BL is spent for growing some virtual radial dissipation scattering speed $v_{diss}$ of the mass $M_{FB} = \text{const}$, i.e. with designations (26) in a view of

$$M_{FB}\frac{dv_{diss}}{dt} \sim \frac{dF_{FB}}{dr_{FB}}\Delta r_{FB} + \frac{d\mathcal{E}_{FB}/dt}{v_{diss}},$$

at perturbation $v_{diss} \to v_{diss} + \Delta v_{diss}$, taking $\Delta v_{diss} = (\Delta v_{diss})_0 \cdot \exp\beta t$, neglecting simultaneously occurring oscillations and indirect (inexplicit) dependence of mass and other parameters BL on time/speed and $r_{FB}$, that is for simplicity at $\Delta r_{FB} \sim 0$, we will obtain in linear on $\Delta v_{diss}$ approximate inequality for the order of value of Langmuir's lifetime $t_{life}$

$$t_{life} \gtrsim \frac{1}{\beta} = \frac{M_{FB}v_{diss}^2}{d\mathcal{E}_{FB}/dt},$$

which agrees with an analogous estimation of lifetime (30) at acceleration BL along a travelling trajectory.

# 10. Conclusion

Fireball is challenging, complex phenomenon naturally unifying a number of diverse, simultaneously present tightly coupled processes with moving in the direction of a very strong electric field some thermodynamically non-equilibrium plasma object.

The proposed Dynamic Dipole Model [8], it seems, is the first model of this kind and has heuristic nature. There is specified the basic opportunity of calculation of basic parameters BL, but full precise calculations are now hardly possible due to the highly integrated complex nature of the problem: need of parameter knowledge of basic acting processes and very painstaking calculation of missing parameters of non-equilibrium plasma with knowledge of the many cross sections of interrelated collision processes and collective problems of setting plasma volume blackness degree. So empirically, various options were specified not contrary to the subjective temperatures $T_m$, $T_e$ and BL luminosities $w$ observations. The main: with these versions of temperatures and luminosities BL can exist and be stable only for a thermodynamically highly non-equilibrium plasma at the least temperatures $T_m$ several thousands K and $T_e \gtrsim (8-12)\times 10^3$ K. In addition, there are relevant observations of BL with very large energies and energy densities up to $\sim 10^{12}$ J/m$^3$ that do not agree with the above energy constraints and DDM results in Tables 5 and 7 and can be due to other causes. Some samples of parameters of low energy BL are represented in Tables 6 and 8.

Existence of BL is possible only at (1) strong enough dipole separation of the positive and negative charges providing supplying BL with energy at its movement. Thus the higher limit of electron density $n_e^{(FB)}$ is determined by atmospheric electric field $E_{env}$ at which the streamer or local high-voltage discharges yet there do not arise, and it can make up to $\sim 10^{16}$ cm$^{-3}$. This corresponds to the usually very small polarizability factor $\gamma$ only up to 1 and a few more (see Table 7 in Section 6).

At the same time, the opposite requirement (2) of superimposing ion and electron clouds which is sufficient as for collisional transfer of the travelling momentum of electrons to particles of BL and for ionization and recombination with light emission, should be fulfilled.

For consistently simultaneously to satisfy both requirements, it is necessary to be $0 < \gamma \lesssim 1$, to which corresponds higher limit $n_e^{(FB)} \lesssim 10^{16}$ cm$^{-3}$ with indefinite lower limit. One of constraints can be natural constraint on highly increased at small $n_e^{(FB)}$ velocity of BL, with additional energy supply to it at keeping temperatures $T_m, T_e$ sufficient for supporting the low gas density in BL and ionization and energies balance, as it follows from Table 8 with minimal electron density $n_e^{(FB)} > 10^{14}$ cm$^{-3}$, $\gamma > 10^{-3}$.

It is possible to assume that in the region of prevailing ionization processes at forward electron edge of BL, the essential contribution to a short-wave light emission spectrum is created by lines of radiation from high excited electron levels (blue luminescence), and at back ion edge of BL there will be essential contribution of continuous spectrum of recombination on high electron levels (red luminescence).

A typical BL is not experimentally reproduced. This corresponds to a large number of often very exotic models of BL with more or less plausible, but more frequently rather questionable assumptions. Therefore, the only criterion of truth BL model so used might be only naturalness of these assumptions (e.g., temperature non-equilibrium $T_i \ll T_e$ of conventional electrical discharge in the rarefied gas and being observed the "cold" BLs) and model qualitative correspondence to the numerous disparate observations of diverse features of BLs.

Let us emphasize that the energy of BL is defined mainly by the balance of energy supplied during its movement in the atmospheric electric field and the losses which are mainly due both to radiation and corresponding to it forward recoil jet creating brake resistance to the translation movement of BL.

With growth of amplitude of Langmuir oscillations, Langmuir-like waves with increasing amplitude can play the prevailing role with excitation Langmuir-like waves. Observations of BL, being full set of problems of non-magnetized low temperature plasmas with a huge variety of observed BL displays and properties, give an occasion to studying non-uniform, thermodynamically non-equilibrium plasma of BL with its inner non-Maxwellian counter flows of charged particles, inhomogeneities and turbulent instabilities, for example, the simplest type of beating $\omega_k$ at summing of Langmuir oscillations or the sum of chaotic Langmuir waves with close frequencies $\omega_1, \omega_k$ and $\omega_k = (\omega_1 - \omega_k)/2$, presumably with volume granular structure $\sim \cos(\omega_k t + \boldsymbol{k} \boldsymbol{r})$ with small $\omega_k$ and very low group velocity of the wave clot $d\omega(\boldsymbol{k})/d\boldsymbol{k}$ (the sometimes observed structure of extended along BL axis "watermelon seeds" (V. L. Bychkov, private note) directly followed by destruction of BL; for studying formation of small local electron avalanches "bristle" from BL surface, accompanied by weak crackling and formation of small peaked plasma ejections; kinematic instabilities connected with accelerating movement of BL which is finishing by its explosion or dissolving in air; coupling energy of destruction at explosion BL at its stopping by an obstacle with addition of surpassing energy of the simultaneous local discharge current of the very strong environment atmospheric electric field surrounding and feeding it; changing BL form with passage through intact glass or small holes, and results of many other observations.

The fundamental mechanism of a fireball is opened in detail with a reconstruction of a complete picture of this phenomenon.

## 11. Afterword

The basic mechanism and the processes proceeding in a fireball and defining its existence are revealed.

1. DDM is based on simple and natural physical principles and agrees with the most part of observations. Dipole nature of BL is supported by the very fact of BL translational motion in atmosphere.

2. Movement of BL as a single whole with correspondingly constant equal translational accelerations and speeds of electrons, ions and neutrals leads to the equation connecting the sizes of BL, electron densities, FB plasma polarizability and tension of atmospheric electric field $E_{env}$. It means that transfer to ions and molecules the excess momentum, which is acquired by BL electrons in electric field $E_{env}$ at possible acceleration of BL, occurs mainly for the account of inelastic electron/neutrals collision processes. Crucial significance when using this equation of balance of the force of atmospheric electric field acting on BL electrons and the forces within BL has a huge friction force of not-electrical nature arising under acceleration BL by external electric field and being tightly "tied" to the bulk of BL the electron cloud ("friction clutch"). But in the uniform motion of BL with constant velocity this frictional force within BL is zero and electrons and all BL are tied mainly with forces of attraction electrons/ions. A huge friction force when accelerating electron cloud at which restructuring can take place (for example, increasing the temperature $T_e$ and correspondingly of electron number $N_e$ with increase of the energy inflow to the fireball preventing its dissipation) can sharply reduce the time of transition to the new steady state of BL without its rupture and dissipation.

   Reduction of the time at which restructuring can take place (for example, increasing the temperature $T_e$ and correspondingly of electron number $N_e$ with increase of the energy inflow to the fireball preventing its dissipation) can lead to a state of the new stability of BL.

3. Defining role in supporting BL play its movement, inflow of energy of an atmospheric electricity and radiation energy loss.

4. BL has considerable excess volumetric positive charge keeping it from scattering, an Coulomb analogue of the surface tension.

5. At movement of BL, backward and the weaker forward reactive recoil jets are forming.

6. Expected small excess of temperature BL in the jet behind BL can lead to convective sometimes observed rotation BL.

7. Very large observed energies at BL destructions and correspondingly exceeding large estimated energy densities of BL could allow at destructive impact of a fireball to trees and other objects unambiguously testify it to be explained presumably by the joining discharge of the atmospheric electricity because BL is supported by the very strong surrounding atmospheric electric field and is its indicator. It can be connected with local increase in electric field tension and change of the mechanism of breakdown at approach BL on small distance to a standing out Earth sources of electric potential, some analogue of power discharge from the tip of lightning rod. Observable post-traces of passage of a huge electric current at destructive impact of a fireball to trees and other objects unambiguously testify it.

8. All BLs can travel with acceleration, small or strong, at least on the initial and final stages; $dv_{FB}/dt \neq 0$, though the acceleration can be small because of the small relative contribution of an inertial term into the general balance of energies of BL (cf. Eq. (22)). Stationary translational movement of BL with constant velocity (without acceleration) is not ruled out.

9. Stability of BL is caused by the high thermodynamic non-equilibrium with typical electron density $n_e^{(FB)} \sim (10^{15} \div 10^{16})$ cm$^{-3}$, $T_m = T_i < T_e \lesssim 12\,000$ K, $p = 1$ at and the non-equilibrium radiation with small blackness degree $\varepsilon$ as it is typical for low pressure electric discharges.

10. Small blackness degrees $\varepsilon$ are meant, according to Kirchhoff's law and numeric spectral calculations [11], the transparency of BLs in visible light (in accordance with observations pointed out in [1]).

11. At expected presence of two free parameters, stability of BL to smooth change of its parameters and a variety of its forms can be defined by the presence of variety of conditional minima and extended long hollows in a two-dimensional surface of parametric space. There is also possibility of degenerate solutions (branches of solutions) for some given $E_{env}$.

12. Table 6 demonstrates that observed lethal defeats of a fireball are not necessarily caused by it as that, but there can be result of influence of the large atmospheric electric field whose indicator it is.

13. Maximal energy densities in DDM are constrained by the values not more than $E_{spec} < (10^8 - 10^9)$ J/m$^3$ .

14. BL DDM traveling occurs in a wide range of speeds, which is consistent with observations. BL travels with maximal velocities at decreasing temperature $T_e$, low value $\gamma E_{env}$ ( but with relatively large $E_{env}$) with the lowered electron density and low energy of BL (see Table 8).

15. A decisive role in defining the parameters of BL plays polarizability degree $\gamma$ (usually $\gamma \ll 1$) of plasma fireball.

16. Decrease of the atmospheric electric field can lead to slow decreasing or fading BL, its explosion (at catastrophically fast disappearance of electric field) or sudden stopping BL before an obstacle, there is not ruled out sometimes possible observed decay of BL into several smaller BLs with smaller $E_{env}$ and less energy expenses to support all set of BLs. But this statement requires analytical or numerical check.

17. The typical BL can produce high frequency sound. Lower key corresponds to low electron densities.

18. Illustrative estimate of BL oscillation frequencies gives some plausible hint to a possibility of a preferably thin, high tonality sound accompanying BL flight. Because of the above-stated possibly accelerated fly of BL, the sound tonality can vary in time: BL sings (David Finkelstein, private message). BL sometimes can really "sing", changing a timbre to in conformity with slowly varying internal and external conditions. But Tables 9 and 10 indicate possibility of the wide range of oscillation frequencies, which demand observational testing. Constraints on maximal admissible velocity of BL according to Table 8 lead to increasing $n_e^{(FB)} \gtrsim 10^{15}$ cm$^{-3}$ with the oscillation shift to higher frequencies. Let's notice that lifetimes at Langmuir-like and kinematical instabilities are of the same order of dozens seconds (see Section 9).

19. The problem of Ball Lightning is very difficult calculation problem of coupled collision and electrostatic closely connected interactions.

20. DDM defines the basic experimental requirements needed for the pilot obtaining BL.

21. It is necessary to remind that the predicted possible imbalance of rates of collision interactions inside BL can lead to expected, inherent to BL, regimes of weak or strong acceleration with corresponding contraction of BL lifetimes.

22. Sometimes observable disappearance of BL without explosion is a hint on possibility of the short-time (short-lived) BLs with low electron densities $n_e^{(FB)} \sim (10^{14} \text{-} 10^{15})$ cm$^{-3}$ and low temperature $T_e$ (Table 8).

23. In presence of the force of atmospheric electric field which causes BL existence and towing, weak air flow can move BL in the transverse direction (across the lines of force of atmospheric electric field). At the same time possibility of displacement BL by wind along the lines of force of electric field is determined by stability level of BL to changes in the rate of supplying BL with energy by its supporting electrical field, and is problematic. It appears nevertheless possible rapid small scale turbulent local changes of direction of electric field. With these limitations the photographed whimsical loop trajectories of BL motion in turbulent thunderstorm flows cannot be considered as the definitive argument, which completely rules out BL DDM. Trembling of hands or concussion at shooting with a long exposition is not excluded also.

24. BL supply with energy is determined by the speed of its movement in electric field. Therefore, for example, an attempt to stop BL by oncoming airflow, if it does not increase its speed, leads to the termination of its existence. Visual motion of BL can not be obtained by usual vector summing of velocity BL in direction of atmospheric electric field and velocity of air counter-flow due to dependence of BL kinematic properties on direction of its motion relative to direction of electric field, that is dependence on the angle between electric force line and the stream direction.

25. In light of the above, it seems that BL can be blown away across its motion (not along! Usually due to a very large tractive force) by the wiff that can sometimes help to rescue himself. Contact with BL facilitates lethal breakdown caused by decreasing ohmic resistance to atmospheric electricity.

26. Preliminary consideration on BL lifetimes $t_{liff}$ leads to expectation of reasonable values $t_{life} \gtrsim 10-20$ s, in good accordance with observations. BL lifetime has the likelihood nature and can reach in quiet atmosphere up to many tens seconds.

27. As follows from Table 6, low energy BLs are theoretically possible with predominance of thermodynamic energy $\sim p V_{BL}$.

28. Due to high temperature of BL and very large effective traction force of the atmospheric electric field $F \sim \gamma |e| N_e E_{env}$ in the tens of pounds or more affecting the fireball, it could soften the window glass in sizes $\sim 2 r_{FB}$, give it a slight convexity of forward shape and squeeze out the round hole (with melted edges?) or break it (private message from V. L. Bychkov) .

$$* * *$$

Let's note, that specific thermodynamic energy $E_{th} \sim pV/V$ equal for 1 at to $1.01 \times 10^5$ J/m$^3$, can be much less than specific electrostatic energy $E_{spec}$ with exception BL with very small electrostatic energies (cf. Tables 5 and 6). But despite it, very strong defeat action of BL can be defined by presence of the accompanying very strong atmospheric electric field $E_{env}$, indicator of which is BL (see Table 6). The density of BL energy is defined mainly by electron density and, at BL sizes $\sim$ (10 - 20) cm (at such, but not the giant BL sizes) and $n_e^{(FB)} \sim 5 \times 10^{15}$ cm$^{-3}$, at electron temperatures $T_e \sim 12\,000$ K can reach up to $\lesssim 10^8$ J/m$^3$ at rather small total BL energy. That is the most importantly, the intensity of the atmospheric electric field $E_{env}$ supporting BL is $\sim (1-10)$ kV/cm and locally even more (!!). Energy of the powerful local discharge of the atmospheric electricity accompanying explosion of BL can exceed energy of BL by several orders of magnitude. Similar synenergetic action could be observed at a simultaneous coupling of BL explosion and the local discharge of atmospheric electricity supporting BL with the observed huge cumulative energy density up to $10^{12}$ J/m$^3$ related to the in reality very small BL volume.

According to section 6, it follows from the law of conservation of BL momentum and energy, that at keeping constant an observable BL speed and its invariable view, radiation with loss of BL energy into external space without direct transfer of BL momentum and energy to environment air should lead to directed forward recoil jet of cold air with the same rate of losses of the energy causing braking resistance of BL movement to traction force of atmospheric electric field. It implies the presence of collision recombination processes with transmission the forward momentum of electrons to arising at recombinations neutral molecules. It is in principle possible precise extremely labor consuming scaled numerical simulation of basic collision processes with the account of internal electric field, however it seems that regularly the more or the less inbalance will be leading to accelerated movement of BL. Achievement of ideal full balance with null acceleration seems therefore to be substantially casual. However the spontaneous self-organizing is also possible with transition to the stable state. It should be noted that at said transfer mechanism of momentum, the energy loss due to radiation and to the resistance to motion should be equal.

The next most important task is to study the conditions of BL emergence associated with arising seed discharge with an indispensable condition for the initial appearance of hot, highly rarefied local plasma cloud with non-equilibrium temperatures $T_e > T_m$, presumably in the field of condensation of atmospheric electric field, for example, on a random wind-borne into the atmosphere debris, at the antinodes of the electromagnetic waves, in discharges near the Earth towering charged objects (trees, arc discharge in trolley contacts, power cables, power corona discharge of metal rods, separated plasma fragments of linear lightning etc.).

The hypothetical appearance of the seed cloud of hot strongly rarefied non-equilibrium plasma can be schematically represented by considering two imaginary plane parallel extended electrodes (anode and cathode), between which, at tensions $E_{env} \lesssim$ few kV/cm and distance $r$, that is not enough for the occurrence of a linear lightning streamer discharge, there is possibly arising of a narrow plasma channel breakdown, for example, with windblown tree wet leaf or other electrically weak conductive random object or some tip between the virtual electrodes. In this case energy is supplied to electrons at a rate $d(|e| E_{env} N_e r)/dt$, $N_e \ll N_m$ where $r$ is distance between electrodes, and partially is transferred from electrons to heavy particles. But the heavy particles expend energy at a rate $\sim p\,dV/dt$ not so much on the thermal conductivity than on the lateral expansion of the originally narrow channel of electrical breakdown, forming a volume seed cloud of BL with temperatures $T_m \ll T_e$. The rarity of observations of BL is due to the random nature of the described specific conditions of BL seed discharge.

The viability of BL DDM is supported by a wide range of possible parameters of DDM BL, probably due to presence of two free parameters, which corresponds to the vast diversity of the observed forms, characteristics and behavior of BL creating a picture of its mystery, unpredictability and too entangled nature. It is also possible degeneration, which leads to two or more branches of BL.

Plasma as highly labile substance is easily accepting incredibly diverse forms as mythological Proteus, it is inherent its nature, but there is nothing inexplicable or mystical in it.

**Acknowledgement.** I express my sincere gratitude to V. L. Bychkov (Head of the Department of Physical Electronics, Moscow State University) for stimulating criticism and an indication of literature. My special thanks and gratitude is to Professor of the Yaroslavl State University A. I. Grigorjev for efficient support and valuable comments.

# *Appendix 1*

## Distinctive features of Dipole Dynamical Model (DDM) of Ball Lightning (BL)[†]

### Abstract

The results of for the first time proposed DDM BL are briefly discussed. The options of DDM BL corresponding to parameter space of temperatures $T_{BL} < T_e$, non-equilibrium parameter $\alpha$, radiation power $w$, tension of atmospheric electric field $E_{env}$ and corresponding them BL sizes, velocities of BL movement along force lines $E_{env}$, electron densities and specific energies of electrostatic dipole charge separation are estimated. BL lifetimes are defined by possibility of growing modified Langmuir oscillations (with sound frequencies occurrence) and BL destruction at its accelerated movement in atmosphere. Characteristic parameters and properties of DDM BL are well agreed with the most part of observations.

In the presented here report the basic ideas of Dipole Dynamical Model of a fireball (DDM BL) are resulted only. More detailed statement and estimations are given in the previous work [1].

DDM [1], [2], [3], [4] is unique, as its existence is caused by energy inflow to BL electrons for the account of its movement along force lines of atmospheric electric field $E_{env}$. Owing to charging asymmetry of gas-plasma dipole, the action of positive atmospheric electric field on ions is compensated by recoil at collisions of heavy ions with molecules of dense cold air in the back part of BL with formation, probably, slightly warmed-up back recoil jet of surrounding atmospheric air [1].

Electrons play a role of the locomotive "towing" behind them all the BL. Earlier such model, apparently, was not considered because of virial theorem interdictions, suitable for the closed systems with weak interaction with external objects limiting specific BL energy by thermodynamic value $pV/V \sim 1.01 \times 10^5$ J/m$^3$ (cf. [5], [6]). Inapplicability of the virial theorem to BL is caused by that BL is strongly interacting part of infinitely extended in time and space system of electric field and atmospheric air in which it moves. The values of potential and kinetic energy of BL and its parameters are defined by balance of inflow of energy at BL movement in electric field and, according to estimations [1, 2], mainly by radiation losses that is not connected in any way with the virial theorem.

Movement of BL as a single whole means that the force of atmospheric electric field operating on electrons setting them in motion, is equal to the force operating on the whole mass of BL. It would seem, it does not correspond to inverse proportionality of accelerations of the whole BL mass and electron mass. Equality of force of ions traction by electrons (and owing to collisions and ions recharge, traction of the whole BL) and forces of traction of electrons by the external electric field is expressed by Eq. (1)

$$f = \frac{m_e}{m_{av}} \frac{e^2 N_i N_e}{(2r_{FB})^2} = |e|\gamma E_{env} N_e, \quad E_{inn} = \gamma E_{env}, \qquad (1)$$

where $m_e$, $e$ are accordingly, mass and charge of electron; $m_{av}$ is average mass of a particle in BL; $r_{BL}$ is effective BL radius; $N_i$, $N_e$ are numbers of ions and electrons in BL; $E_{inn}$ is dipole field tension inside BL; the polarization factor $\sim 10^{-2} \lesssim \gamma \lesssim 1$ is of an order of ratio of effective force of the external electric field acting separately on electron or an ion in BL, to average force of Coulomb interaction of the neighboring charges (replacement $E_{env}$ for $\gamma E_{env}$ must be made in all expressions in the works [1, 2]). At forces of traction of ions (and all BL) by electrons and traction of electrons by an atmospheric field, their accelerations should be equal owing to fast electron momentum collisional redistribution in BL.

Justification of Eq.(1), following from observable integrity of BL, should be that force of traction BL by electrons is defined by interaction of overlapping clouds of ions, electrons and neutral molecules in which charge distribution and their interaction at acceleration BL are defined by transfer of acquired by electrons momentum $|e|\gamma E_{env} N_e t$ to the whole BL, including electrons, as a result of mainly inelastic collisions of electrons with neutral molecules, ionization and recombination, recharge at collisions of ions and neutrals, only partly by Coulomb analogue of a surface tension caused by BL considerable excess volumetric positive charge owing to lateral ambipolar diffusion [1], and in very small degree by Coulomb interaction of ion and electron clouds according to (1). It prevents scattering of BL at sharp increase $E_{env}$.

The condition of equal accelerations at different almost on five orders of value masses of BL and electrons is paradoxically combined with equal forces of an electrostatic attraction of ions and electrons and $|e|\gamma E_{env} N_e$, operating on these masses.

Options of BL with various variants of the initial parameters $r_{BL}$, $E_{env}$ chosen with the account of observations BL, to temperatures $T \lesssim 12\,000$ K, $p = 1$ at are calculated. Instability of thermodynamically equilibrium variants [1], and also their very large radiation powers inappropriate to observations are shown. Calculation of nonequilibrium options $T_e$, $T_i = T_m$ with introduction of a new free non-equilibrium parameter $\alpha$ with non-equilibrium real electron density $n_e^{(FB)} = n_e/\alpha$ and observable power of radiation $w = W/\alpha^2$, are fulfilled where $n_e$ and $W$ are correspondingly equilibrium values of electron density and of radiation powers usually $\sim (100-200)$ W at temperature $T_e \sim (6000-12\,000)$ K at pressure 1 at with subsequent recalculation $w$ and $n_e^{(FB)}$ to temperature of ions and molecules $T_m \sim (2000-4000)$ K. It is shown only partly similarity of BL to some kind of a direct current discharge of low pressure.

Presence of stability conditions of BL with a minimum of the potential energy due to two free parameters caused by the presence, for example, $w$ and $T_e$ or $\alpha$ is shown. All other BL parameters at given $E_{env}$ can be calculated in principle from Eq. (1), stability condition, and balance equations of conservation electron numbers and the energy, with the need for the difficult calculations. For reasonably admissible initial parameters of BL which are not contradicting to observations, options of non-equilibrium BL are calculated at any way setting $E_{env}$, $w$, $T_m$, $T_e$, some of which are resulted below in tables (1 - 3) [1]. Speeds of BLs, electrostatic energy of charges separation $E_{full}$, density of electrostatic energy $E_{spec}$ and electron densities (which are constraint by comparison with the data of observations BL usually to the range $n_e^{(FB)} \sim (10^{14}-10^{16})$ cm$^{-3}$) are calculated.

Because of a number of restrictions including reasonable radiation power and tension of the streamer or local high voltage breakdown $E_{env} \lesssim (10-20)$ kV/cm, densities of electrostatic energy of small and large BL do not exceed $E_{spec} \lesssim (10^8-10^9)$ J/m$^3$. Estimated in some observations $E_{spec}$ up to $\sim 10^{12}$ J/m$^3$ can be consequence of wrongly attributed to BL accompanying energy of the high-voltage breakdown whose trigger BL can be usually under tension $\sim 1$ kV/cm and locally much more tension of atmospheric electric field supporting its existence. With this field, whose indicator it is, there can be connected also lethal effect of a fireball. Not BL kills as that, but atmospheric electricity accompanying it. Breakdown is facilitated by contact with BL with decreasing ohmic resistance (for example, with lethal defeat of an animal or the person). The mechanism of the attached atmospheric breakdown at fireball impact into coming out objects (a tree, etc.) can be some analogue of the kind of breakdown in atmosphere from a lightning rod edge.

Owing to the extremely small mass of BL, its trajectory, sometimes especially whimsical in an environment of closely located diverse objects or owing to atmospheric turbulence, continuously follows small-scale changes of directions of local electric field with the smallest effort for moving BL across (not along) force lines of an electric field along which it moves. Thus BL can move along force lines only with the electron edge of a dipole directed forward into a positive direction of electric field, paradoxical way even at excess positive charge of BL.

The mechanism of gasdynamic braking BL by formation of a forward brake jet of recoil owing to recombination processes in BL is opened. Radiation carries away only energy, but not momentum. Thus, electrical energy received at movement BL is spent approximately for radiation with the formation of a brake recoil jet caused by it at forward electron edge of BL (i.e. brake force of uniformly moving BL at creating a forward recoil jet with its power, approximately equals the traction force and is $F_{br}(v_{FB}) = 2w/v_{FB} = |e|\gamma E_{env}N_e$ with the doubled velocity).

There is estimated Boltzmann smoothing to several millimeters of distribution function of charges at forward and back edges of BL in atmospheric electric field.

At the fixed atmospheric electric tension

$$E_{env} \sim |e| \left[ n_e^{(FB)} \right]^{2/3} = \text{const} \,[1] \qquad (2)$$

polarization parameter $\gamma$ depends on BL radius that at $E_{env} = \text{const}$ can cause the modified Langmuir oscillations in the field of high sound frequencies lower and above the range of sound audibility $\sim (4 - 7)$ kHz. In [1] the illustrative variant of calculation, using parameters of BL from Table 6, of the modified Langmuir oscillations with frequency 3.4 kHz is resulted at $n_e^{(FB)} = 1.1 \times 10^{15}$ cm$^{-3}$ according to the formula

$$\nu_{osc} \geq \left[ \frac{7}{12\pi} \frac{m_e}{m_{av}^2} \frac{n_e^{(FB)}}{n_m^2} \frac{E_{env}^3}{|e|} \right]^{1/2}, \qquad (3)$$

where $n_m$ is particle density in BL. Thus, possibility (depending mainly on electron density) of high key sound changing its tonality (because of possible acceleration BL with change of its parameters) owing to BL oscillations arising for the account of excitation of modified Langmuir oscillations of plasma BL (BL "sings" on David Finkelstein's expression in private message) are predicted. The whistling sound can be created also by the back recoil jet. Probably, sometimes visible granules in BL [7] grow out of turbulent Langmuir heterogeneity, in particular of instability before disintegration of BL.

The balance of energy of moving BL is defined by the obvious equation

$$|e|\gamma E_{env}N_e(t)v_{BL} - w(t) - W_{diff}\,t - W_{gd}(t) - W_{th}(t) - \frac{d}{dt}\frac{M_{BL}(t) \cdot v_{BL}^2}{2} = 0, \qquad (4)$$

where $N_e = n_e^{(FB)}V_{BL}$; $V_{BL}$ is BL volume; force of atmospheric field operating on BL is $F = |e|\gamma E_{env}N_e(t)$; $w(t)$ is radiation power; $W_{diff}$ is the rate of diffusion energy losses; $W_{gd}(t)$ is power spent for overcoming gasdynamic resistance to BL movement including power on creation the forward recoil jet; $W_{th}(t)$ is power of thermal losses; last term of Eq. (4) accounts for energy losses on overcoming forces of inertia. Stationary traveling of BL corresponds to $dv_{BL}/dt = 0$.

Parameters of BL are defined by Eq. (1), balance equations (or the rates of processes) and stability conditions at two free parameters of BL. Thus, difficult to achieve problem is the finding dependences of BL: $r_{BL}(t)$, $T_e(t)$, $\alpha(t)$, $T_{BL}(t)$, $n_e^{(FB)}(t)$, $w(t)$, $W_{diff}(t)$, $W_{gd}(t)$, $W_{th}(t)$ on time, corresponding to the varying energy $\sim |e| \gamma(t) E_{env} N_e(t) \cdot 2r_{BL}(t)$, with the subsequent solving of the differential equation (4) for speed $dv_{BL}(t)/dt \neq 0$ where as the first approach it is represented reasonable to use perturbation theory with $dv_{BL}(t)/dt = 0$.

Thus, asymptotically divergent solutions are possible with explosion or with gradual disappearance of BL. The presence of the more or less acceleration rates of BL according to Eq. (4) with $dv_{FB}(t)/dt \neq 0$ and possibility of occurrence of kinematic and Langmuir instability may be usual. All BLs can move with acceleration, at, least, on initial and, possibly, final states with instability occurrence. Acceleration can be a sign and possible internally inherent cause, besides changes of an atmospheric environment and changes $E_{env}$, defining BL lifetime, reaching according to observations to tens seconds. However to find analytical or numerically simulated dependences of these interconnected macroscopic and microscopic parameters on time is extremely difficult, therefore the most perspective thing appears to calculate various toy variants with the entry conditions close to the solution with $dv_{BL}/dt \sim 0$ with asymptotical limit $dv_{BL}(t)/dt \to 0$ (if it exists), and then comparing results with observations (cf. [1]). Estimations made in [1] show that BL lifetime has likelihood character depending on atmospheric conditions, and can reach many tens seconds in agree with observations [7].

DDM BL is relatively cold with temperature $T_{BL} \sim (1000-4000)$ K, but enough hot for observed at radius $\sim (5 -- 10)$ cm burning holes in window glasses. It has usually rather weak luminosity of $\sim 100$ W that specifies its nonequilibrium character with $T_{BL} \equiv T_m = T_i \ll T_e$.

Observable passage BL through intact window glasses can testify about strong polarization of DDM BL.

BL can be observed without visible connection with a linear lightning.

Strictly speaking, it would be possible to expect oval form of BL, characteristic for a dipole, with non-uniform internal structure. The spherical form is reasonable idealization of the first approach. The large excess volumetric positive charge of BL keeping electrons from running away plays a role of analogue of the surface tension, explaining variability of BL forms and its possibility to get through cracks and narrow holes.

Owing to very small calculated spectral and integral blackness degree $\varepsilon$ of BL according to data presented in the literature, it is transparent in visible region of the spectrum.

The calculated velocities of movement of BL [1] vary in a wide range, increasing with reduction of radius of BL.

Vertical "falling" BL sometimes observed from the sky agrees with representation about the positive charge of the Earth and confirms presence of negative charge on the forward edge of dipole BL.

DDM BL predicts explosion or destruction of BL at its stopping by any obstacle.

Disappearing BL observed sometimes without explosion specifies possibility of short-time BL with low electron densities $10^{15} > n_e^{(BL)} \gtrsim 10^{14}$ cm$^{-3}$ at low density of BL with still enough high temperatures $T_m$, $T_e$ (see Table 8).

Destruction of BL or its collapse can occur at very fast change $E_{env}$ with which have not time relaxing collision processes.

Presence of two free parameters of BL leads to existence of two and more branches of the solution and a variety of kinds of BL with a random walk over the degenerate states [2].

In the presence of force of the atmospheric electric field causing existence and towage of BL, possibility of moving BL by wind along force lines of a field is defined by the level of stability BL to changes of the rate of supplying energy to BL and is problematic. Nevertheless, fast small-scale local turbulent changes of a direction of electric field are possible. With these restrictions, a photo of freakish loops of BL trajectory in turbulent storm streams cannot be considered as the decisive argument deleting existence DDM BL.

In a storm cloud, velocity of apparent motion of BL is not defined by the vector sum of BL velocity relative air and velocity of an air stream, including with possibility of zero total velocity stopping BL at nonzero speed of a stream. Stopping BL stops its existence because of the termination of its energy feed by atmospheric electric field. "Blowing away" BL can sometimes help rescue from it.

Moving gas-plasma dipole is not directly an analogue of the static glow discharge confined in a finite space with virtual cathode and anode. In view of the chaotic thermal motion of electrons in the fireball a balance of ionizations and recombinations is setting without the usual electric current between the virtual cathode and the anode (or rather weak one).

One can suppose that ionization by accelerated electrons prevails at the electron front of BL, recombination prevails in the back of BL, which accumulate the ions. However owing to integrity BL and a smoothing background of prevailing rarefied neutral gas component of BL, it is very difficult to image presence of sharp heterogeneity of BL. At the same time due to asymmetry of BL dipole, weak volumetric microflows of electrons and ions can arise between dipole ends which can appreciably disturb Maxwellian distribution.

It is necessary to notice that DDM BL is qualitatively agreed with the most part of various observations and explains them, what eliminates mystical character of BL nature. Nevertheless, apparently, some observations mentioned in very detailed review [7], do not give in to a simple explanation within the frame of DDM BL. For example, defeats of people by penetrating radiation, disappearance next to the skin metal objects (next to the skin metallic adornments, necklaces, rings to the finger, bracelets) at defeat by a fireball, covering person with a shining cloud of a fireball without harm for him and a number of others.

In work [1] are presented tables of a considerable number of variants of calculated parameters of BL. It allows to determinate conditions of possible experimental creation BL: it is necessary to create extended electric field with the electric tension on the threshold of occurrence of streamer or local high voltage breakdown (what is a difficult engineering problem) and seed strongly non-equilibrium local discharge, probably by means of focused high power microwave discharge.

DDM BL has heuristic character of a principal substantiation of model and requires the further completion and development. For example, estimations of BL lifetime owing to kinematical and Langmuir's instabilities (many tens seconds, see [1]) and more fundamental detailed calculations of parameters of BL are required.

Plasma is extremely unstable mobile substance easily accepting improbably various appearances as mythological Proteus, this is the nature inherent in it, and in this rare and really exotic phenomenon of BL, there is not present anything inexplicable and mystical.

(Further it is followed with above Tables 6, 7 and 8 of basic text).

_______________________________________

### Additional notes

**1.** Estimation of parameters of nonequilibrium plasma with introduction of nonequilibrium parameter $\alpha$ requires essential development. According to Kirchhoff's law in thermodynamically equilibrium plasma and neglecting strongly reabsorbed spectral lines structure of atom spectra, degree of blackness $\varepsilon$ is proportional to degree of absorption $1 - e^{-\tau}$, where $\tau$ is optical thickness $\tau \sim \sigma_{abs} n_m^* r_{FB}$; $\sigma_{abs}$ is absorption cross-section, $n_m^* \propto n_e$ is density of absorbing atoms and molecules excited on high levels with distribution of higher states according to temperature $T_e$. After transition to thermodynamically nonequilibrium plasma, $\tau$ is proportional to $(T_e/T_m) \cdot (n_e/\alpha)$, and corrections to $\varepsilon$ and $w \sim W_r/\alpha^2$ are defined by the factor:

$$\xi \sim \left. 1 - e^{-\tau \cdot (T_e/T_m)/\alpha} \middle/ (1 - e^{-\tau}) \right. \text{ at } n_e^{(FB)} \sim n_e/\alpha$$

(see Section 3). Thus, we have $w \sim \xi W_r/\alpha_1^2$, $n_e^{(FB)} \sim n_e/\alpha_2$ with some different values $\alpha_1$ and $\alpha_2$ (cf. Section 4). Therefore the error of calculations at the large optical thickness with $r_{FB} \sim (100-1000)$ cm in Table 5 can make to several tens percent only in an intermediate part of the interval $\varepsilon \ll 1$, $\varepsilon \lesssim 1$ and keeps at least the general trend of changing the properties BL at transition to BLs with very large sizes (if those exist).

**2.** In Section 6 earlier used re-designation $E_{env}$ to $E_{eff}$ is interpreted as designation of an effective field of polarization on the scale of $\sim (1-100)$ V/cm at the external atmospheric electric field equal to Coulomb field between neighboring charges in scales $\sim (1-10)$ kV/cm. In this connection it is offered to re-designate (rename) in Sections 2 -- 5 and further small effective $E_{env}$ to $E_{eff} = \gamma E_{env}$, $\gamma < 1$ and to keeping appropriate designation $E_{env}$ for the much more atmospheric electric field. Expression for polarization parameter $\gamma$ is thus resulted (see Section 6 and also Eq. 26). In Section 3 stability conditions as presence of a minimum of potential energy $U$ are discussed

$$\frac{d}{dT_e}U(r_{FB}, w, \gamma E_{env}, T_e) = 0$$

with values $(r_{FB}, w, \gamma E_{env}) = const$ at variable $T_e$. The obtained result corresponds with atmospheric field $E_{env}$ of the order value $\sim (1-10)$ kV/cm. In case of $(r_{FB}, w, \gamma E_{env}) = const$ it is expected that similar calculations using tabular data $n_e$ and $\varepsilon(r_{FB}, T_e)$ in equations

$$\frac{dU}{dT_e} = \frac{d}{dT_e}(|e| \cdot \gamma E_{env} \cdot 2r_{FB} \cdot V_{FB} \cdot \frac{n_e}{\alpha}) = \frac{d}{dT_e}\left\{ B \cdot \frac{n_e^3(T_e)}{\left[\varepsilon(r_{FB}, T_e) \cdot T_e^4\right]^{3/2}} \right\} = 0,$$

$$B \equiv \frac{\pi^{1/2}}{9} e^2 r_{FB} (\frac{w}{\sigma})^{3/2} = const, \quad (r_{FB}, w, E_{env}) = const,$$

where $\sigma$ is Stefan-Boltzmann constant, $V_{FB}$ is BL volume, should lead to close similar results of Section 3 with performance of stability conditions in wide parameter space of two free parameters of BL.

The mystical much variety of BL forms is explained by that at five parameters which define BL: $r_{FB}$, $E_{env}$, $T_e$, $T_m$, $w$, only three connecting them equations are available: integrity of BL as a single whole; stability; balance of energies; conservation of the momentum of movement. Remaining two parameters have likelihood nature and are arbitrary (free). The same whole parameter space of stable BLs should not depend on any other concrete choice of two independent free parameters.

# *Appendix 2*
# Ad initio principles of the Dipole Dynamical Model of Ball Lightning

V. N. Soshnikov

Plasma Physics Dept.,

All-Russian Institute of Scientific and Technical Information
of the Russian Academy of Sciences
(VINITI, Usievitcha 20, 125315 Moscow, Russia)

## Abstract

There are exposed principles of the dipole dynamical model of ball lightning (DDM BL) which was proposed recently and studied in detail for the first time by the author [3]. BL moves along the force lines of atmospheric electric field with their slight incline compensating Archimedean BL buoyance force. Longitudinal asymmetry of dipole BL with the recoil force of rear jet of cold (or slightly heated) air which compensates force of acting atmospheric electrical field on ions, leads to BL steady movement in atmosphere with the energy pumping of traveling forward electrons by atmospheric field that, in turn, explains the very long BL lifetimes. Stability of ball lightning is due to the presence of two free parameters of BL as it is thermodynamically non-equilibrium plasma. Energy balance is achieved with influx of energy from the atmospheric electric field and radiation losses and related with them losses of resistance to fireball movement in atmosphere. To this we must also add the crucial condition of integrity BL as a single whole due to previously obtained required (but insufficient!) balance condition of the acting on electrons force of atmospheric electric field and interior forces of BL.

Ball lightning (BL), as a mysterious manifestation of atmospheric electricity, has generated a lot of hypothetical explanations, as a rule, only for some of its observable separate characteristics. Results of numerous sporadic observations and some of the most developed models of BL, sometimes with an exotic character, are set out detailed in a recent comprehensive review [1]. At the same time, I proposed a new dipole dynamic model of BL (DDM BL) based on the observed features of BL and natural relations that characterize the thermodynamically non-equilibrium rarefied weakly ionized air plasma ($T_m \sim 10^3$ K $\ll T_e \sim 10^4$ K), which qualitatively explains almost all of the observed features of BL in their entirety [2 - 6] with the most detailed and conclusive exposition in [3]. In the following are presented the fundamental principles of the model, confirmed by estimates and calculations given in [3].

Fireball is challenging, complex phenomenon, unifying a number of diverse, simultaneously present tightly coupled processes of moving BL in the direction of a very strong electric field the some non-equilibrium plasma object.

Any model must explain and quantify the properties of BL, the most frequently observed in the aggregate: BL sizes of about 5 -- 20 cm; BL speed, often of the order of a few $m/s$; related to BL energies of explosion with energy densities up to $\sim 10^{12} \, J/m^3$ [1]; rounded form, sometimes transforming into a "snake", with penetration through the cracks and intact window panes; the BL lifetime up to many tens of seconds; moderate brightness, such as a 100-watt lamp; a relatively low temperature of about $(1000\text{-}3000) \, K$, sometimes enough to cause a burn or to melt the glass pane; BL behavior at the end of its life: the disappearance or explosion.

1. The dipole model BL can exist only at the observed forward movement of BL under the influence of the atmospheric electric field. The driving force of BL translational motion is electrons which are affected by the effective force $f$ of the external electric field $f = |e| N_e E_{eff} \equiv |e| N_e \gamma E_{env}$. With a uniform translational motion of BL electrons entrain ions with the same force, and with them the whole mass of BL consisting of a rarefied weakly ionized plasma of charged and neutral particles of air at a pressure of $\sim 1$ at. External electric force $-|e| N_e E_{eff}$ acting on the ions is almost completely compensated by the recoil jet produced by collisions of hot ions with molecules of cold dense air behind the BL. It is also possible accelerated BL motion in a certain range at which electrons gain some additional momentum from the external electric field which is then redistributed to the whole BL due to predominantly inelastic collisions of electrons with molecules and ions while maintaining integrity of BL as a whole. Condition of BL movement as an integer with a predominance of transport of acquired by electrons momentum to BL by inelastic and elastic collisions is the equation [3] (see above more detailed in Section 3, pp. 5 – 6):

$$\gamma E_{env} = \frac{\pi}{3} \frac{m_e}{m_{av}} |e| \frac{r_{FB}}{n_m} n_e n_i, \qquad (1)$$

where $E_{env}$ is atmospheric electric field, $\gamma \ll 1$ is polarization factor, $m_e$ is electron mass, $m_{av}$ is average mass of the particle in BL, $n_m$ is BL particles concentration, $r_{FB}$ is BL radius, $n_e \sim n_i \ll n_m$ are electron and ion densities. This equation is valid for a uniform and slightly accelerated motion of BL. Equation (1) can be also clearly and easily understood as the necessary (but not sufficient!) condition of existing BL as an integer whole.

The observed integrity of BL can only be explained by the fact that at BL movement with the same effective acting forces of atmospheric field on the electrons $|e| E_{eff}$ and the dipole interaction electrons/ions, the coupling BL with electrons ("friction clutch") at uniform motion and small acceleration is ensured by the necessary to equalize the accelerations with $M_{BL}/M_e$ times larger BL non electric pushing by very specific friction force of mainly inelastic collisions of electrons with molecules of BL.

Atmospheric electric field tracks with the force $f_{tr} = |e| \gamma E_{env} N_e$ electron cloud which is tightly tied to BL in its movement in BL by the arising friction clutch force $f_{tr} \cdot (M_{BL}/M_e)$ (see also in more details above Sect. 3, pp. 5 – 6).

The "friction clutch" is the main force which determines the acceleration of BL as a whole under the influence of the atmospheric field acting on its electrons, including also some small force addition due to increasing the temperature $T_e$ of the electron cloud, respectively, with small increasing degree of ionization and acceleration of BL with the possible transition to a state of the new stability of BL.

Crucial for DDM BL is an elementary equation of equality dipole attraction of electrons and ions of BL (per unit mass of BL) and the impact force of the atmospheric electric field on the electrons (per unit mass of the electron cloud) to ensure the completion of BL energy with its traveling movement as a whole, including a constant BL speed when driving in a medium with resistance acting on BL as a whole with balancing traction force.

Transmission of total electron momentum to a huge (as $M_{BL}/M_e \sim 10^5$) mass of BL is accompanied with very large rate of this process with an estimated very small time $\tau$ of momentum transfer in multiple collisions

$$\tau \sim \frac{m_m}{m_e} \cdot \frac{1}{n_e \bar{v}_e} \cdot \frac{1}{\sigma_e(T_e) N_e}, \qquad (2)$$

where $m_m$ is average mass of BL particle, $n_m$ is their density, $\bar{v}_e \sim \sqrt{kT_e/m_e}$ is electron velocity, $\sigma(T_e)$ is momentum transfer effective cross section of inelastic (and elastic) electron-molecule collisions in BL. At the same time there is growing of $T_e$ and polarization factor $\gamma$ and attraction of $E_{env}$. This expression determines some critical acceleration with the destruction of BL as $a_{crit} \tau^2/2 \ll r_{BL}$, where $a_{crit}$ is maximum permissible acceleration. Nonetheless, analysis of kinematical instability of BL with the ability of its "rupture" remains, apart from the acceleration/deceleration instability, an open problem.

Similar transfer of excess momentum of electrons to BL mass occurs at uniform motion of BL with constant speed and is spent on compensation for the loss of energy in the air resistance to its motion. However friction force is very specific, and if there is no acceleration, the electron cloud within BL is immovable relative to it, and the friction force is equal to zero.

2. Because of polarization of the plasma fireball as collective phenomenon, effective field acting on the charged particles can be estimated approximately as $\gamma E_{env}$ where $E_{env}$ is real atmospheric electric field reaching several kV/cm, but less than the field tension of a lightning streamer breakdown. Polarization coefficient $\gamma$ is determined by the critical ratio of the impact force on the electron $|e| E_{env}$, and the Coulomb interaction between neighboring BL electrons of the order $e^2 n_e^{2/3}$, i.e. $\gamma < E_{eff}/|e| n_e^{2/3}$. Thus, values $E_{env}$ used in [2] ought to be renamed to $E_{eff}$ and to confront them much more greater values of the external atmospheric electric field $E_{eff}/\gamma$, which in [2] are not given, keeping all other results. Typical values [3] are: $0.01 \lesssim \gamma < 1$; $E_{eff}$ in the range from about 10 V/cm to several hundreds V/cm, and $E_{env}$ up to ~ 10 kV/cm.

3. Due to the ambipolar diffusion, BL has a significant positive charge [3], what does not prevent its forward movement in atmosphere (due to the effect of ion momentum compensatory jet behind BL), but creating the effect of a large surface tension. At sufficiently high electron temperature at which the ionization rate is much greater than the loss rate of ambipolar diffusion, the positive charge of BL provides its integrity. BL always travels with its negative charge directed forward.

4. DDM BL is not possible without account for substantial thermodynamic non-equilibrium $T_e \gg T_i = T_m$ of plasma fireball. This was accounted by introduction of the free non-equilibrium parameter $\alpha$ with which the equilibrium radiation at a temperature $T_e$ and pressure 1at (using the estimated blackness degrees $\varepsilon$ at the base of hemispherical volume [7]) was decreased in $\alpha^2$ times, while reducing approximately the density of electrons in $\alpha$ times. Parameter $\alpha$ and the size of BL (or, equivalently, the size $r_{BL}$ and $E_{eff}$ or $E_{env}$, either equivalently any other pair combinations of parameters) can be regarded as two independent free parameters of BL which ensure its stability.

5. The presence of two free parameters leads to a natural presence of local minima, for example, for the radius of BL in its dependence on the temperature $T_e$ that corresponds also to the potential energy minimum depending on the temperature, or equivalently in other similar pair combinations determining the stability of BL. Such minima do not exist in the case of the plasma fireball thermodynamic equilibrium, which in this case has moreover impossible enormous energy radiation [2, 3].

6. The energy $|e| N_e E_{eff} v_{BL} t$ that fireball gets at its movement in atmospheric electric field is spent mainly to the radiation (with the account for non-equilibrium parameter $\alpha$) and on the resistance to its movement in atmosphere. However, loss of the energy by radiation, it would seem, does not create a direct mechanical momentum of resistance to movement of BL. The one of the mechanisms for the creation of the momentum loss is electron-ion recombination with transfer of the electron momentum to forming neutral molecules, creating a counter recoil jet in air in front of the electronic part of BL . Thus, the radiation carries away half of the incoming energy, and the other half is spent in overcoming the resistance of counter recoil momentum resulting from electron momentum transfer to cold dense air molecules in the front edge of BL, that makes up half of the recoil momentum $f \cdot \Delta t$ opposing to forward momentum of BL. The balance of these energies and BL train force determine the wide range of calculated speeds of BL movement (see additional consideration in ***Appendix 3***).

7. In the framework of DDM, parameters of various BLs were calculated in the form of tables that are estimative in nature (toy calculations) demonstrating the possibility of simultaneous good agreement with numerous observations [3].

8. It should be noted that the estimated maximum energy density in DDM BL does not exceed, respectively, thus the value given in [1] up to ~ $10^{12}$ J/m³ can be attributed to the impact of BL as a trigger of large-scale local atmospheric discharge collecting electricity from a large volume of thundery electrified air. In this case, BL is an indicator of its supporting large electric fields in surrounding atmosphere, but BL itself has relatively limited resource of energy.

9. BL can be accompanied with low-frequency noise (generated by an air recoil jet behind BL) and high-frequency sound (generated by an analogue of the Langmuir oscillations of electrons in moving BL [3]).

10. BL lifetime reaching many tens seconds is determined by inhomogeneity of the atmosphere, fluctuations of the atmospheric electric field and acceleration instability of BL. Within the framework of DDM the existence of immovable BL for any long time is not possible, but one can still imagine the plasma cloud in the static focused strong high-frequency (microwave) electric field with the energy supply by microwave radiation and plasma wave processes.

11. BL requires for its emergence the seed discharge creating a hot cloud of non-equilibrium plasma with low density. Let imagine two hypothetical extended (flat or linear) parallel virtual electrodes between which the electric field is of the order of a few kV/cm, which causes a breakdown on accident windblown object (wet leaf of a tree, etc.) or small roughness or sharp on the electrode to form narrow plasma channel. We can assume that the energy imparted to electrons by the electric field will be transported to heavy atoms and molecules, and the latter will not spend so much energy by heat conduction, as for the across expansion, which will result in a volumetric highly rarefied moderately hot plasma cloud to form BL mainly near objects at the Earth surface. BL is a fairly rare phenomenon because of the incidental occurrence of a seed discharge.

Continuously ongoing virtual "collapse" of BL (recombination) is offset by an opposite polarization retain of ion and electron clouds by an external electric field and the simultaneous recovery processes of ionization at high enough (~ 10 000 - 12 000 K) electron temperatures. The "dissipation" of BL is prevented by attraction of electrons with the excess positive BL charge and at the same time with filling up ions by ionization which compensates the loss of ions as in the ambipolar diffusion and recombination. BL existence is due to balance of these processes (plus the stability condition in the presence of two free parameters) with the energy supplying by the atmospheric electric field at BL movement in the positive direction (the electron part of the dipole forward) along the electric field lines and at the same time inseparable from BL recoil jet of air behind generated by the energy of the atmospheric electric field expended in equal amounts to creating the backward jet and the summed energy loss of radiation and resistance to BL movement. This mechanism makes the DDM BL very natural and consistent at the same time with the most part of available observations. Region of the BL existence parameter space is constrained by the calculation of the many BL toy variants. The above estimates and toy calculations set out principles of DDM BL within which it ought to perform future improved calculations to create a complete theoretical model which includes the entire set of multiple processes responsible for existence of BL.

In this paper we has not considered the completely different widespread cluster model of ball lightning [8], [9] that binds a fireball with a very complicated fine-cellular structure of clusters of burning or excited impurities produced during a lightning impact to a particular object. It is assumed that the cluster forms a rigid frame which supports the shape of a fireball with a very high energy content of the surface electric charge.

This model seems to have a visible confirmation in observations the behavior of the tiny "ball lightning" produced under specific electrical discharges between metal electrodes, in electric/magnetic field. The model has a very special nature and may explain, in our opinion, with all its plausibility, only special forms of ball lightning.

The main problem of such (static) models is the presence of an energy reserve (obtained from linear lightning?) which is required to explain the observed BL lifetime up to $\gtrsim 100$ s. Only movement along the lines of the atmospheric electric field gives life to a fireball and is inseparable from it.

Multicomponent combination of plasma dynamical properties that characterize moving plasmoid of DDM BL plasma clot is different from the huge number of existing models explaining selectively only some BL features not related to the rest of the observed features of a real fireball. Thus, the theoretical model of BL should include a composed set of several closely related plasma processes with two free parameters, and such diversity of well known usual natural plasma processes does not appear an artificial and far-fetched conglomeration.

DDM BL can not be considered without analyzing BL as the indissoluble community of equivalently important underlying coupled processes which makes its many-component nature and leads to a natural explanation of the most of its rather diverse observed properties. DDM BL is quite different from the existing difficult, sometimes many-component exotic construction models explaining usually only small part of the observed properties of BL without proper explaining the main feature: the observed long BL lifetime.

**Note 1.** Translation of the paper submitted to the Journal of Technical Physics (Russian Federation):  Сошников В. Н. // "Дипольная динамическая модель шаровой молнии".  Журнал Технической Физики, РФ, 2013. In February 2014 an article was rejected without giving any reason. This ***Appendix*** is also planned as a possible short paper for English-language Journal.

A new revisit of all the presented material was made with correcting small inaccuracies in text without changing the results of calculations and any principled positions.

**Note 2.** Most of this Section was published: Soshnikov V. N., "Ad initial principles of the Dipole Dynamical Model of Ball Lightning". International Journal of Mathematical and Theoretical Physics, 2014, May, v.4, n.3, pp. 84 – 87.

**Afterword:** **About numerous models of Ball Lightning.** I think, that the previously expressed my first view on existing BL models is confirmed.

**1.** All current BL models are based on the presence of internal or borrowed from any external object the energy for BL (chemical reactions, borrowing from the linear lightning, electrical discharge in a stationary fixed localization of thickening force lines of the atmospheric electric field, an**d** so on), to explain a long BL lifetime for which existence ones are forced to use the assumptions of a wide variety, the most exotic and incredibly sophisticated sources of energy: with unusual electrodynamical processes, multilayer sources, including shells of microparticles and water vapor/droplets, MHD vortices, electrical microcapacitors, structured combustible clusters of silicon or metal atoms, high-energy nuclear particles with the quantum-mechanical properties, and many other contrived usually difficult constructions which are related to the field of professional activities of the authors. Thus, in these models, there are naively used household phantoms: (1) the presence of a rigid (solid or liquid) the outer envelope of BL; (2) the presence of a hard inner sceleton (or core), defining BL external form (cluster theory); (3) the availability of powerful initial stock of consumable energy which determines the observed long lifetime of the fireball. But really, all this is not!

**2.** Here, for example, is given illustrative offhand Internet listing of some small part of the current works from the huge entire list with characteristic titles/abstracts:

**A model for Ball Lightning.**
David Fryberger, Stanford Linear Accelerator Center, Stanford University, Stanford, California 94309. See: http://www.slac.stanford.edu/pubs/slacpubs/6250/slac-pub-6473.pdf

**ABSTRACT**. A model for ball lightning (BL) is described. It is based upon the vorton model for elementary particles, which exploits the symmetry between electricity and magnetism. The core, or driving engine, of BL in this model is comprised of a vorton-antivorton plasma. The energy of BL, which derives from nucleon decay catalyzed by this plasma, leads, through various mechanisms, to BL luminosity as well as to other BL features. It is argued that this model could also be a suitable explanation for other luminous phenomena, such as the unidentiØed atmospheric light phenomena seen at Hessdalen. It is predicted that BL and similar atmospheric luminous phenomena should manifest certain features unique to this model, which would be observable with suitable instrumentation. SLAC PUB 6473. October 1994. *Invited talk presented at the First International Workshop on the UnidentiØed. Atmospheric Light Phenomena in Hessdalen, Hessdalen, Norway, March 23-27, 1994.* [Extravagant known theory of electric-magnetic duality of Maxwell equations with the existence as ordinary particles (e, p, n, μ, ν, etc.) and their magnetic analogs, and hypothetical vortex pair associations in elementary nucleonic particles vorton-antivorton, forming the core of BL with energies near to the thermonuclear ones]

**The Physical Theory of Ball Lightning.**
S.G. Fedosin, A.S. Kim. English version of the paper: Applied physics (Russian Journal), No. 1, 2001, pp. 69 – 87. Bukireva Str, 15, Perm State University, 614990 Perm, Russia. E-mail: intelli@list.ru.

**ABSTRACT.** The analysis of modern models of ball lightning displays, that they are unsatisfactory on a series of tests. The mode of ball lightning is offered, which exterior electronic envelope is retained by interior volumetric positive charge. The compounded electron motion in an outer envelope creates the strong magnetic field driving a state of ionized hot air inside ball lightning. The conditions of origin surveyed, the estimates of parameters of ball lightning of different power are made. [Two shells with electrostatic-magnetic retention (confinement)]

**Magnetically dominated plasma models of Ball Lightning.**
Pekka Janhunen. Finnish Meteorological Institute, Geophysics Department, 21.12. 1989.

**ABSTRACT.** Recently, ball lightning models based on MHD force balance equation have been proposed. An upper bound for the magnetic energy for these models is presented. The possibility of weakly ionized plasma models is considered with estimates on lifetime and energy content. The lifetime is found to be too short, if the electron-neutral and ion-neutral effective collision frequencies behave in the usual way. The possibilities to get around these restrictions are briefly analyzed. [MHD vortex]

**New model and estimation of the danger of ball lightning.**
M. L. Shmatov. J. Plasma Physics (2003), vol. 69, part 6, pp. 507–527. Ioffe Physical Technical Institute, 194021 St. Petersburg, Russia.

**ABSTRACT.** A new model of ball lightning is proposed. The main model assumption is that ball lightning has a core consisting of clouds of electrons and totally ionized ions which oscillate with respect to each other. According to the model, ball lightning emits high energy photons that are sometimes dangerous for human beings, and in a number of situations it can kill humans by electric pulses; the ball lightning energy can be of the order of 106 J and even greater. The electric charges that need to be injected into the atmosphere to create ball lightning and the currents, providing the injection of such charges, are estimated. These estimates predict that ball lightning can be created in the experiments with ordinary lightning or powerful electrical installations. [Electrostatically oscillating ion-electron shells]

**Theoretical Studies of Long Lived Plasma Structures.**
Maxim Dvornikov. http://www.researchgate.net/publication/45907047. Theoretical Studies of Long Lived Plasma Structures.
**ABSTRACT.** We construct the model of a long lived plasma structure based on spherically symmetric oscillations of electrons in plasma. Oscillations of electrons are studied in frames of both classical and quantum approaches. We obtain the density profile of electrons and the dispersion relations for these oscillations. The differences between classical and quantum approaches are discussed. Then we study the interaction between electrons participating in spherically symmetric oscillations. We find that this interaction can be attractive and electrons can form bound states. The applications of the obtained results to the theory of natural plasmoids are considered. Comment: 12 pages in pdf, 1 jpeg figure; the contribution to the proceedings of the II International Conference "Atmosphere, Ionosphere, Safety" (AIS-2010), Kaliningrad, Russia, June 21-27, 2010. [Classic and quantum-mechanical ion-electron oscillations of spherical two-shell two-component plasma ball]

**Anatomy of a Lightning Ball.**
Peter Weiss. http://www.phschool.com/science/science_news/articles/anatomy_of_lightning.
**ABSTRACT.** The notion that aerosols may be a part of ball lightning goes back to at least the 1970s, but it's currently winning unprecedented attention. One way out of the conundrum is to add features of an aerosol to a plasma theory of ball lightning. An aerosol's additional material can form a structure, host long-lasting chemical reactions, store electric charges, and otherwise account for observed ball-lightning properties, An aerosol's additional material can form a structure, host long-lasting chemical reactions, store electric charges, and otherwise account for observed ball-lightning properties, Turner and others argue. [Plasma-chemical cluster nature]

**Solid charged-core model of ball lightning.**
D. B. Muldrew. Ann. Geophys., 28, 223–232. - 2010.
**ABSTRACT.** In this study, ball lightning (BL) is assumed to have a solid, positively-charged core. According to this underlying assumption, the core is surrounded by a thin electron layer with a charge nearly equal in magnitude to that of the core. A vacuum exists between the core and the electron layer containing an intense electromagnetic (EM) field which is reflected and guided by the electron layer. The microwave EM field applies a ponderomotive force (radiation pressure) to the electrons preventing them from falling into the core. The energetic electrons ionize the air next to the electron layer forming a neutral plasma layer. [Two-shell structure with its retention (confinement) by intermediate microwave guide]

**Ball Lightning Study.**
Eric W. Davis. 24 February r– May 2003. Final report. Air Force Research Laboratory.
http://www.foia.af.mil/shared/media/document/AFD-091008-049.pdf
**ABSTRACT.** Nachamkin model; microwave; plasmoid resonance; axially symmetric; force-free; atmospheric maser caviton; electromagnetic vortex plasmoids. The focus of this study was to review and analyze the axially symmetric force-free time-harmonic plasmoid model developed by Nachamkin (1992) for a previous Air Force Research Laboratory study. Three alternative ball lightning concepts similar to axially symmetric force-free time-harmonic plasmoids were identified and evaluated for their experimental potential, and are described in the report in detail as proposed experiments. The first new concept is the atmospheric maser caviton, the second concept is based on electromagnetic vortex plasmoids generated by micro-discharge devices and sustained by quantum vacuum energy, and the third concept is a [classified] program of the Air Force funded in the 1950s-60s. [Atmospheric maser or fireball as a moving focused microwave discharge (soliton) in atmospheric microwave channel]

**The Ball Lightning Puzzle.**
John G. Cramer. http://www.analogsf.com/0512/altview.shtml
**The Alternate View.** This leaves the theories, involving some form of electromagnetic process. I find the most compelling of these to be the maser-caviton theory of Handel and Leitner, building on previous ideas of Kapitza. The basic idea is that the high electric field pulse accompanying a lightning stroke in a flat terrain can create a population inversion from the storage of energy in the rotational energy levels of water molecules. The large atmospheric maser (i.e., laser for microwaves) thereby created can occupy a volume of several cubic miles and can last for many seconds. This restless sea of stored energy can form an elaborate and irregular standing wave pattern, which "spikes" in some locations. At such a spike, the ball lightning discharge forms and is powered by the action of the maser, drawing energy from the entire maser volume. The result is what is called a "soliton" of electromagnetic radiation, forming a hot cavity in the high-field region surrounded by a glowing plasma of ionized air.

  I like the Handel-Leitner theory because of its scientific fiction possibilities. One could imagine a weapon that harnesses the energy present in a thunderstorm to throw lightning balls at the opposition. [Atmospheric aerosol maser] …

**Ball Lightning, Wikipedia.**
An extensive history list with the brief review of observations and theories of BL until 2013, including even such exotic solid-state theory of BL, BL as a black hole or a nuclear antimatter meteor, or neutrino-antineutrino annihilation etc is presented, for example, in Wikipedia, see: http://en.wikipedia.org/wiki/Ball_lightning .

  **3.** To this it can be added as an internal source of energy the neutral Rydberg atoms, plasma oscillations, plasma toroidal waveguide with non-damping light of laser beam, BL as specific phase state of dense cold (300 K) plasma, and much more, until the existence of undiscovered powerful quantum-mechanical forces. All these hypotheses are clearly one-sided nature. It is shocking that among many tens of numerous ideas and models of BL, there is not detected at least a single one in which it was assumed to be an extremely simple, extremely natural robust idea in which energy supply of BL would occur, likewise in accelerators of charged particles, from the BL movement in environmental energy ocean of atmospheric electricity with the automatically following indefinitely long BL lifetime and excellent compliance with numerous observations of BL and the rich diversity of its properties.

In this connection, it should also be again noted observation of the strong recoil jet behind BL (A. I. Nikitin, see page 11, item 12, Ref. [2]) in favor of DDM BL.

**4.** Experimental probing DDM BL might be the development of small artificial seed discharges and even hot pyrotechnic flares in the thunderstorm atmosphere. For example, in the great review of V. L. Bychkov *et al.* (see the basic text, Ref. 1, p.24; there are also described several new modified models of BL), it is pointed out the emergence of BL from electric discharge of suddenly disconnected from electric network trolley contact.

One can compare this with arising fireball on the tip of the lightning rod and its "falling" to Earth.

**5.** The value of this work which was beginning by me back in 2009 and systematically continued and regularly replenished until November 2014, is a fundamental justification of completely new, which apparently has by nobody previously offered, the most natural and consistent with observations, the dipole dynamic model of ball lightning, which, of course, needs to be promising in the further detailed elaboration of its diverse and specific units and, may be, possible further experimental and observational confirmation activities.

## *Appendix 3*

## Dipole Dynamical Model of Ball Lightning (DDMBL) (status and brief overview)

### V. N. Soshnikov

Plasma Physics Dept.,

All-Russian Institute of Scientific and Technical Information
of the Russian Academy of Sciences
(VINITI, Usievitcha 20, 125315 Moscow, Russia)

A new theoretical model of ball lightning (DDMBL) has been presented in materials of the two previous Conferences on Electrical Phenomena in the Atmosphere (Yaroslavl State University, 1-4 July 2011 and 4-8 July 2013). The most complete presentation of DDMBL is represented by me at the work: *V. N. Soshnikov.* " Comments to support the Dipole Dynamical Model (DDM) of Ball Lightning (BL)", ArXiv.org/physics/gen-ph/arXiv:1007.4377. This presentation has repeatedly supplemented with parts which are published in the Journal "Электронная обработка материалов", 2012, v.**48**, n. 6, pp. 54 – 64, and in International Journal of Mathematical and Theoretical Physics, 2014, v. **4**, n. 3, pp. 84 – 87. The publication of this model of ball lightning in February 2013 (in Russian version) was denied without giving any reason by Editory of the Journal of Applied Physics (RF) (sent to the Russian "Журнал Технической Физики").

As far as I know, this is the only model of such kind, since it is not based on the initial, but in principle not sufficient reserves of energy consumed by a fireball and leading respectively to its lifetime significantly lower than the observed values.

First DDMBL is based on the continuous replenishment of its energy due to the motion of the dipole BL in atmospheric thunderstorm electric field in analogy with the energy of a charged particle moving in an electric field of the accelerator. The electric field imparts energy to electrons, while the opposite effect of the atmospheric field force acting on ions is balanced by recoil of heavy ions with a mass equal to the mass of the molecules of the surrounding cold air with the much more density with the formation of the recoil jet of cold air. This model was first made possible after calculations of the emissivities of hot air at temperatures up to about 10 000 K and pressure 1 at (*L. M. Biberman et al.*, 1962 – 1964 ("Optical properties of hot air", M. Nauka, RAS USSR, in Russian), pp. 320 (1970)) in which I calculated the emissivity contribution of diatomic molecules and NO$_2$).

Crucial for DDMBL is an elementary equation of equality of dipole attraction between electrons and ions of BL (per unit mass of BL) and the value of the force of the atmospheric electric field acting on the electrons (per unit mass of the electron cloud) to ensure the completion of energy BL with its translational movement including a constant speed when driving in an environment air with resistance acting on BL as a whole, with balancing traction. Stability of BL is determined by the appearance of the second free parameter, which is the degree of temperature non-equilibrium (electron and ion temperatures ratio), the usual for low temperature plasma of electrical discharges of low pressure (low density). Radiation is the main source of energy losses (neglecting ambipolar diffusion losses at enough high temperatures) and causes loss of energy for the resistance to BL movement in the surrounding air equivalent to radiation losses with transfer of the electron translational momentum and energy to molecules. Because of the large charge-exchange cross sections in collisions of ions with molecules, fast electron momentum transfer to molecules and the excess BL positive charge, it is homogeneous gas-plasma analogue of "suspension of asymmetrically distributed charges in neutral liquid " moving as a single whole (but with possibility of the limited movement of charges in the liquid).

Uniform motion of BL means that the momentum imparted to electrons by the atmospheric electric field, respectively, to the whole BL, is balanced by the momentum of resistance to movement of BL.

But radiation of BL carries away only energy, but does not carry away and not create momentum.

Due to the inelastic and elastic collisions, electrons transfer their excess momentum to BL molecules, collisions of which with the molecules of the cold dense air at the forefront of BL determine the momentum of resistance to BL movement equal to the electric field train momentum, and the electron-electron collisions result in the quasi-Maxwell velocity distribution in the swarm of electrons which leads to energy loss for radiation corresponding to the electron temperature $T_e$.

To compensate for the radiation energy losses there is needed inflow of the energy power $\left|eE_{eff}N_e\right|v_{BL}$, where $v_{BL}$ is some constant speed of existing BL movement. But the law of conservation of momentum at the same speed $v_{BL}$ requires expense of exactly the same energy to overcome the resistance force to BL movement.

Both the laws: energy and momentum conservation, are consistent with a doubling of the speed value $v_{BL}$ with equal energy expenses of the radiation and the resistance to movement of BL.

Since the implementation of these relationships requires consistency of very complex collisional micro-processes in the fireball, it is more common really accelerated movement of BL than is usually subjectively perceived by witnesses the movement with the constant speed.

. In this case, as a first approximation is introduced also polarizing factor $\gamma$ binding atmospheric field with the so-called effective field acting on the formally unconnected together electrons and unconnected together ions of BL. The factor $\gamma$ determines the ratio of the force of atmospheric electric field to single electron and the force of electrostatic interaction between two electrons at an average distance between them in BL.This approach requires further clarification, which in the simplest case reduces to multiplication the polarization coefficient $\gamma$ by some universal constant number (expected value $\lesssim 1$), with possible further consideration dependence $\gamma$ on the configuration, size and density of the electron cloud. This is the heuristic nature of DDMBL, which is confirmed by the natural and reasonable assumptions and is commonly in good agreement with the whole set of observations.

This heuristic model is original, in principle different from many exotic models of stationary BL with the expenditure of energy derived from a linear lightning discharge or from other sources. The lifetime of DDMBL is determined only by the possibility of internal plasma instabilities and changes in the external environment and the labile electric fields of the thunderstorm atmosphere including the possible instability of the accelerated motion of BL, and can reach many tens seconds in accordance with the observations.

Performed multivariate simulation calculations ("toy calculations") indicate the possibility of agreeing DDM at reasonable natural regions of the parameter space of initial data with almost all very diverse disparate observations of ball lightning, up to compliance with emitted by BL sounds.

It is assumed that BL does not have any formative solid, liquid, or any force magneto-electro-dynamic outer shell (skin). BL form is due to BL excessive positive charge (due to the ambipolar diffusion), holding the heavy hot molecules of rarefied weakly ionized plasma within BL due to preventing their diffusion the recoils in collisions with molecules of the cold, dense air) and due to the repeated recharging at ion-molecule collisions (with the ion fraction $\lesssim 0.01$ ) which is equivalent to the time averaged distribution of shares of the total positive BL charge among all molecules. Neutral molecules may also be held within moving BL partly by transferring them the translational momentum of electron cloud by non-elastic collisions and recombination. Of course, there is a small leak of neutral molecules.

It should be noted that the amazing mystical properties of BL can be explained without resorting to electromagnetic, relativistic and other phenomena from other physical areas, but namely by the simultaneous collection of the well-known properties of the alone, but now interrelated in BL, low pressure electrical discharges with non-equilibrium weakly ionized plasma.

Estimated from some existing observations implausible huge BL energy density up to $\sim 10^{12}\,\mathrm{J/m^3}$ can of course be explained by the fact that BL is only indicator of the required for its existence atmospheric fields up to $\sim 1-10\ \mathrm{kV/cm}$, but below the threshold of streamer breakdown (or an usual breakdown at small discharge gaps). Therefore, to the fireball electron here can join a powerful breakdown discharge from the release of enormous electrical energy of surrounding atmosphere. Maximum calculated DDMBL energy density is not more than $\sim 10^9\,\mathrm{J/m^3}$ and decreases with increasing BL size.

BL arises often above ground surface, moves horizontally along the force lines of atmospheric electric field and can be accompanied by an explosion of great destruction power. Additional protection of particularly dangerous areas, such as with highly flammable oil tanks, in addition to the vertical lightning rods, can serve a network of ground electrical current-conducting metal fittings connecting all the electrical grounds (zeros), distributed on the protected area and removing the difference of landscape potentials between them, what makes impossible BL arising and movement on the protected area. This fact was pointed out by the member of Laser Association (RF) Mikhail Vyacheslavovitch Zakharov, whom the author expresses his sincere gratitude.

Experimental creation and observation of BL in atmosphere is only possible when you create a very strong electric field near the threshold of streamer breakdown of the order of thousands $\mathrm{V/cm}$ on the extended trajectories when creating the initial seed discharge to form a hot rarefied gas plasma cloud. To get an experimental analogue of ball lightning one can try to use an easily ionized gas that increases the light emissivity, with low decreased pressure, respectively with reduced voltage of streamer breakdown, which may, however, increase with a decrease in the discharge gap. The problem reduces to the possibility of bringing into translational motion the standing static discharge (perhaps the focused microwave discharge) by the field less than a few $\mathrm{kV/cm}$ with a sufficiently high (up to $\sim(1000-3000\ \mathrm{K})$) temperature rarefaction of gas. Orderly movement of BL (not associated with the wind) is the main diagnostic feature of DDMBL.

It is surprising that almost all recognize the independent movement of BL, but nobody clearly explained how and why this happens!

Extensive demolitions by BL can be caused by BL accompanied very strong atmospheric electric field, which causes (not BL itself!) the greatest damage. Thus the diagnostic BL features can be a very slow speed of BL movement (corresponding to the high temperature and the degree of ionization of BL). But the speed decreasing may also be due to a branch of smooth decreasing the electric field and disappearance of BL without explosion.

An extensive history list with the brief review of observations and theories of BL until 2013, including even such exotic solid-state theory of BL, BL as a black hole or a nuclear antimatter meteor, or neutrino-antineutrino annihilation, BL is gasdynamic soliton with its specific equation etc, is presented, for example, in Wikipedia, see: http://en.wikipedia.org/wiki/Ball_lightning . But until 2014 there is no mention about any models of dipole-dynamic type.

The proposed DD-model of ball lightning is based on simple common fundamental laws of physics (conservation of energy and momentum, BL as an observable moving in atmospheric electric field a single whole object) and does not use any exotic usually complicated hypothetical microstructure elements within BL, and also without, as a rule, constructing of the really not existing BL shell.

BL consists of hot diluted neutral molecules ($\sim 10^{18}\,\mathrm{cm}^{-3}$); the charges account for less than 0.01 of the particles of BL ($\lesssim 10^{16}\,\mathrm{cm}^{-3}$). However, collisions with ions with recharge of molecules and their periodic ionization and recombination in collisions with electrons can be regarded as an equivalent of some average over time weak positive charge over all the molecules, retention them from their escape out of BL without the formation of a clearly expressed the outer limiting shell. It also has significance the reflection of the hot molecules inside BL from the surrounding dense cold air outside BL. Of course, this question: is BL spherical or oval in shape, respectively to dipole asymmetry, requires special analysis (evaluation). The absence of any outer shell may be accompanied by mild forms of rapid variability of BL under the influence of external conditions: rapid change in the atmospheric electric field and other changes in the environment.

Neutral molecules periodically change to the state of molecular ion and vice verse.

For example, let $r_{BL} \sim 5$ cm to be BL radius; $n_m \sim 10^{18}\,\mathrm{cm}^{-3}$ is density of neutral molecules of BL; $w$ eV/s is radiation power of BL equal to $\sim 100$ W; $I_{ion}$ is ionization potential $\sim 15$ eV; $n_i = n_e \sim 10^{16}\,\mathrm{cm}^{-3}$ are densities of ions and electrons; BL volume is $V = 5.22 \cdot 10^2\ \mathrm{cm}^3$, $n_m V \equiv N_m$. Then the time density of the charge-change transition of each molecule per second in its process of ionization→recombination and vice versa will be $\sim 2w / I_{ion} N_m \sim 2.4$ (mol·s)$^{-1}$ at unchangeable $n_m, n_i, n_e$. If the average molecular ion recombination time in BL is $\tau_{rec}$, then the transmitted to ion additional momentum which retains in BL the forming neutral molecule will be $1.2 \times |e| E_{ion} w \tau_{rec} / (I_{ion} \cdot N_m)$, where $E_{ion} \approx \gamma E_{env}$. This value defined by the cross-section of capture of free electron by the molecular ion and stay time of neutral molecule in BL, along with recharging collisions of molecular ion and a neutral molecule and retention of neutral molecules inside the hot fireball when they collide with the dense cold air environment, defines the ability to hold the cloud of neutral molecules in maintaining BL as a single whole with a somewhat blurred BL contour. It should be noted however, that the actual rate of gas ionization is determined by direct and intermediate steps of excitation, as well as, respectively, recombination with electron capture by molecular ions.

In the above proposed serendipitous case the neutral molecules are completely equivalent to one another in attributing to their behavior some properties of particles with a small positive charge, and also to some extent the lack of the property of quasi-neutrality of the considered BL dipole plasma at interaction of its positive and negative charges.

It should be noted however that the actual rate of gas ionization is determined by direct and multiple intermediate steps of excitation, as well as, respectively, recombination with electron capture by molecular ions, but this does not alter the above general considerations. It follows from the balance as a whole of ions, neutral molecules and energy radiation.

The proposed DDM BL contains nothing mystical, it naturally combines logically related process chain in close analogy with the processes of well-studied electrical discharge in rarefied weakly ionized plasmas. It seems surprising that the plasma consists almost of the neutral rarefied air molecules of a low density ($n_m \sim 10^{18}$ cm$^{-3}$, $T_m \sim (3000 - 4000)$ K), and negligible content $\lesssim 1\%$ of electrons and molecular ions ($n_e \sim (10^{15} \div 10^{16})$ cm$^{-3}$; $T_e \lesssim 12\,000$ K), which, however, have a decisive effect on the properties of BL. It is amazing the very simple self-organization of BL when there are two free parameters which does not require any specific type shells of MHD layers, inner and outer shells, presence of clusters Si or Fe creating rigid form of BL, and any other exotic rigid or elastic elements. The phenomenon of Ball Lightning is a single indivisible whole of Fire Ball and the environment medium. This is the uniqueness of the given model, in which the very existence of ball lightning is caused by its movement in ambient atmospheric air and the atmospheric electric field.